\documentclass[prd,aps,showpacs,nofootinbib,showkeywords,eqsecnum,preprint]{revtex4-2}
\usepackage{graphicx,color,amsmath,amsxtra}
\usepackage{epsf}
\usepackage{amssymb}
\usepackage{enumerate}
\usepackage{hhline}
\usepackage{array}
\usepackage{tabularx}
\usepackage[unicode]{hyperref}
\usepackage{graphicx}                
\usepackage{epstopdf}
\usepackage{subfig}
\usepackage{epsfig}

\begin{document}
\pagestyle{myheadings}
\title{ Anisotropic power-law inflation for the S\'aez-Ballester theory non-minimally coupled to a vector field}
\author{Tuan Q. Do }
\email{tuan.doquoc@phenikaa-uni.edu.vn}
\affiliation{Phenikaa Institute for Advanced Study, Phenikaa University, Hanoi 12116, Vietnam}
\author{Phung V. Dong }
\email{dong.phungvan@phenikaa-uni.edu.vn}
\affiliation{Phenikaa Institute for Advanced Study, Phenikaa University, Hanoi 12116, Vietnam}
\author{Duy H. Nguyen}
\email{duy.nguyenhoang@phenikaa-uni.edu.vn}
\affiliation{Phenikaa Institute for Advanced Study, Phenikaa University, Hanoi 12116, Vietnam}
\author{J. K. Singh }
\email{jksingh@nsut.ac.in}
\affiliation{Department of Mathematics, Netaji Subhas University of Technology, New Delhi 110078, India}
\date{\today} 
\begin{abstract}
In this paper, we would like to examine whether the S\'aez-Ballester theory admits stable and attractive Bianchi type I inflationary solutions in the presence of a non-minimal coupling between scalar and vector fields such as $f^2(\phi)F_{\mu\nu}F^{\mu\nu}$. As a result, such a solution will be shown to exist within this theory for a suitable setup of fields.  Interestingly, the considered S\'aez-Ballester theory can be shown to be equivalent to the standard scalar-vector theory via a suitable field redefinition. This means that the obtained solution can be reduced to that derived in an original anisotropic inflation model proposed by Kanno, Soda, and Watanabe. Consequently, the corresponding tensor-to-scalar ratio of this solution turns out to be higher than the latest observational value of the Planck satellite (Planck 2018) due to the fact that $c_s$, the corresponding speed of sound of scalar perturbations of the S\'aez-Ballester  theory, turns out to be one.  This result indicates an important hint that the speed of sound, $c_s$, could play an important role in making the corresponding non-canonical anisotropic inflation cosmologically viable in the light of the Planck 2018 data. To be more specific, we will point out that any modifications of the S\'aez-Ballester theory having $c_s \sim 0.1$ will have a great potential to be highly consistent with the Planck 2018 data. For heuristic reasons, a simple modified version of the  S\'aez-Ballester theory will be proposed as a specific demonstration. As a result, we will show that this modified model admits an anisotropic power-law inflationary solution as expected.
\end{abstract}
\maketitle
\newpage
\section{Introduction} \label{intro}
Recently, the validity of standard cosmologies has been questioned extensively due to the emergence of a number of observational anomalies, which seem to not align with the cosmological principle, the underlying assumption, by which our universe is homogeneous and isotropic on large scales as uniquely described by the well-known Friedmann-Lemaitre-Robertson-Walker (FLRW) metric, e.g., see Refs. \cite{Schwarz:2015cma,Buchert:2015wwr,Perivolaropoulos:2021jda,Abdalla:2022yfr,Aluri:2022hzs} for the most  recent relevant reviews on this issue. For the early universe, two remarkable anomalies  should be mentioned are the cold spot and hemispheric asymmetry of the cosmic microwave background radiation (CMB), which were detected by the Wilkinson Microwave Anisotropy Probe (WMAP) \cite{WMAP:2012nax} then confirmed by the Planck satellite \cite{Planck:2019evm}. See also Ref. \cite{Kester:2023qmm} for a recent independent confirmation of the CMB hemispheric asymmetry using the latest data of the Planck. As a result, the existence of these anomalies might provide an important hint that the early universe might not be isotropic at all \cite{Schwarz:2015cma,Jones:2023ncn}. In other words, an anisotropic universe in the early time might be a reasonable approach to explain the origin of the CMB anomalies \cite{Schwarz:2015cma}. For the late time universe, some other anomalies such as the Hubble tension might also provide more evidences for a breakdown in FLRW cosmology as  pointed out by interesting analysis in Refs. \cite{Krishnan:2021dyb,Krishnan:2021jmh}. It turns out that these analysis follow an important result  announced previously in Refs. \cite{Colin:2019opb,Secrest:2020has} that the present universe might be spatially anisotropic. See also Ref. \cite{Rameez:2024xsn} for  a recent discussion on this issue and Ref. \cite{Boubel:2024cmh} for a recent independent testing of anisotropic Hubble expansion.

So far, we have listed some of the most recent relevant evidences for the possibility that our universe might not be isotropic on large scales. One might therefore claim that the cosmological principle might just be a good approximation, by which the complexity of calculations is reduced significantly, rather than a solid physical assumption. It seems that an anisotropic universe might be a better scenario. If so, ones would ask if any existing physics and/or cosmological theories/models support this possibility. In addition, a further related question would be addressed is that how to connect two anisotropic states of our observable universe, one in the early time and the other in the late time, to each other. Theoretically, such anisotropic universe seems to be in tension with the prediction of the so-called cosmic no-hair conjecture proposed  long time ago in Refs. \cite{Gibbons:1977mu,Hawking:1981fz} by Hawking et al. As a result,  this conjecture implies that all late time states of our universe would no longer be inhomogeneous and/or anisotropic on large scales, i.e., they must obey the cosmological principle, regardless of initial states or conditions. Remarkably, this conjecture is still a conjecture because of the fact that there has been no any complete, general proof for it since the first seminal (partial) proof by Wald for Bianchi spacetimes, which are homogeneous but anisotropic, in the presence of a positive cosmological constant $\Lambda$ \cite{Wald:1983ky,Barrow:1987ia,Mijic:1987bq,Kitada:1991ih,Maleknejad:2012as,Kleban:2016sqm,East:2015ggf,Carroll:2017kjo}. Interestingly, some counterexamples to the cosmic no-hair conjecture have been proposed in a number of cosmological models, e.g., see Refs. \cite{Kaloper:1991rw,Barrow:2005qv,Watanabe:2009ct,Kanno:2010nr} for an incomplete list of literature. Besides, the validity of the cosmic no-hair conjecture has been examined in other scenarios \cite{Tahara:2018orv,Starobinsky:2019xdp,Galeev:2021xit,Nojiri:2022idp}. It is worth noting that some people have pointed out that the cosmic no-hair conjecture would only be valid locally, i.e., inside the future event horizon \cite{Starobinsky:1982mr,Muller:1989rp,Barrow:1984zz,Jensen:1986nf,SteinSchabes:1986sy}.

Remarkably, the unavoidable observational anomalies mentioned above indicate an important consequence that the cosmic no-hair conjecture might potentially be broken down during an inflationary phase of the early universe. If so, counterexample(s) to the cosmic no-hair conjecture would exist during this inflationary phase. It is noted that most of counterexamples have been associated with homogeneous but anisotropic inflationary universes, whose background metrics are of the so-called Bianchi spacetimes \cite{Ellis:1968vb,Pitrou:2008gk,Gumrukcuoglu:2007bx}. In the standard picture of modern cosmology, the inflationary phase is assumed to happen right after the Big Bang and last in a very short period of time. During this phase, the universe expands very rapidly such that the flatness, horizon and magnetic monopole problems can be resolved accordingly \cite{Starobinsky:1980te,Guth:1980zm,Linde:1981mu}. Furthermore, cosmic inflation can give us a reasonable mechanism for explaining the origin of large-scale structures of our present universe \cite{Lyth:2009zz}. Remarkably, many theoretical predictions based on the cosmic inflationary paradigm have been shown to be well fitted by the Planck 2018 data \cite{Planck:2018vyg,Planck:2018jri}. For an up-to-date review of cosmic inflation, one can see Ref. \cite{Odintsov:2023weg}. A huge number of models of inflaton, a scalar field assumed to be responsible for the cosmic inflation, can be seen in an interesting paper \cite{Martin:2013tda}.

Among the claimed counterexamples in Refs. \cite{Kaloper:1991rw,Barrow:2005qv,Watanabe:2009ct,Kanno:2010nr}, only that derived in a supergravity motivated model by Kanno, Soda, and Watanabe (KSW) in Refs. \cite{Watanabe:2009ct,Kanno:2010nr} has passed the stability test, while the others \cite{Kaloper:1991rw,Barrow:2005qv} have been shown to be unstable against field perturbations \cite{Kao:2009zzb,Chang:2011zzb}. As a result, the KSW model has been shown to admit a stable and attractive Bianchi type I inflationary solution having small spatial anisotropies \cite{Watanabe:2009ct,Kanno:2010nr}.  This is indeed beyond the prediction of the Hawking cosmic no-hair conjecture. Consequently, the state of universe at the end of  inflationary phase would still be anisotropic with small spatial anisotropies, whose imprints would be detected in the CMB map. Indeed, the imprints of such anisotropic inflation have been investigated in details in Refs. \cite{Watanabe:2010fh,Dulaney:2010sq,Gumrukcuoglu:2010yc,Watanabe:2010bu,Bartolo:2012sd,Ohashi:2013mka,Ohashi:2013qba,Chen:2014eua}. Other aspects of the KSW anisotropic inflation can be found in interesting reviews \cite{Soda:2012zm,Maleknejad:2012fw}.

As a result, the most important aspect of the KSW anisotropic inflation model \cite{Watanabe:2009ct,Kanno:2010nr}  is due to the existence of a non-minimal coupling between the scalar (a.k.a inflaton) $\phi$ and vector (a.k.a. one-form or electromagnetic) $A_\mu$ fields such as $f^2(\phi)F_{\mu\nu}F^{\mu\nu}$. In particular, this coupling does play the leading role in maintaining stable spatial anisotropies of Bianchi type I background metric during the inflationary phase. It appears that the existence of non-constant function $f(\phi)$ will prevent the vector $A_\mu$ from a rapid dilution during the inflationary phase. Consequently, the corresponding stable hairs (a.k.a. spatial anisotropies) having small values will exist, in contrast to the prediction of the cosmic no-hair conjecture. In follow-up papers \cite{Do:2011zz,Ohashi:2013pca,Holland:2017cza,Nguyen:2021emx,Do:2016ofi,Do:2020hjf,Pham:2023evo}, we and the other people  have investigated  whether the scalar field $\phi$ could affect on the stability of such anisotropic inflation by replacing its canonical form by non-canonical forms proposed in some well-known inflationary models such as the Dirac-Born-Infeld (DBI) \cite{Silverstein:2003hf,Alishahiha:2004eh} and {\it k}-inflation \cite{Armendariz-Picon:1999hyi,Garriga:1999vw}. An interesting result has been obtained in all these papers \cite{Do:2011zz,Ohashi:2013pca,Holland:2017cza,Nguyen:2021emx,Do:2016ofi,Do:2020hjf,Pham:2023evo} is that the cosmic no-hair conjecture is always violated, no matter non-canonical forms of scalar field $\phi$ are. In other words, these non-canonical extensions always admit stable anisotropic inflationary solutions. In Refs. \cite{Do:2016ofi,Do:2020ler,Do:2020hjf,Pham:2023evo}, we have investigated the corresponding CMB imprints of these non-canonical anisotropic inflation. Another interesting result derived in these works is that the studied non-canonical extensions of the KSW model can give rise their tensor-to-scalar ratio highly consistent with the Planck 2018 data \cite{Do:2020hjf,Pham:2023evo}. All these results motivate us to extend our analysis to other non-canonical scalar fields discussed extensively recently. One of them we are currently interested in comes from the so-called S\'aez-Ballester (SB) theory \cite{Saez:1986dil,Saez:1987st} (see also Refs. \cite{Singh:1991vou,Singh,Socorro:2009pt,Rasouli:2017glb,Rasouli:2022hnp,Quiros:2022vhm,Rasouli:2019axn,Rasouli:2022tjn,Singh:2024zvm} for an incomplete list of relevant references). As a result, we will investigate whether the SB theory admits a stable anisotropic inflationary solution in the presence of non-minimal coupling $f^2(\phi)F_{\mu\nu}F^{\mu\nu}$. In case such a solution is confirmed to exist, we will study its corresponding tensor-to-scalar ratio to see if it is viable or not in the light of the Planck 2018 data. 

This paper will be organized as follows: (i) A brief introduction of our study has been written in Sec. \ref{intro}. (ii) A basic model setup of the S\'aez-Ballester theory will be presented in Sec. \ref{sec2}. (iii) Anisotropic power-law inflationary solutions of the  S\'aez-Ballester theory will be solved in Sec. \ref{sec3}. (iv) Then, the stability and attractive property of the obtained anisotropic inflationary solutions will be investigated in details in Sec. \ref{sec4}. (v) An equivalence between the SB and standard scalar-tensor theories will be shown in Sec. \ref{sec5}. (vi) In order to see the cosmological viability of the obtained inflationary solutions, the corresponding tensor-to-scalar ratio will be worked out and compared to the Planck 2018 data in Sec. \ref{sec6}. (vii) A modified SB model along with its anisotropic power-law inflationary solution will be presented in Sec. \ref{sec7}. (viii) Finally, our conclusions and further remarks will be given in Sec. \ref{final}. Some additional calculations will be listed in the Appendix \ref{appendix1}.
\section{Model setup} \label{sec2}
 First of all, a general action of non-canonical KSW models is described as follows \cite{Ohashi:2013pca,Do:2020ler},
 \begin{equation} \label{general-action}
S = \int {d^4 } x\sqrt {- g} \left[ {\frac{{R}}
{2}  +P(\phi,X)  - \frac{1}
{4}f^2 \left(\phi\right)F_{\mu \nu } F^{\mu \nu } } \right],
\end{equation}
where the reduced Planck mass, $M_p$, has been set as one for convenience. In this action, $F_{\mu \nu }  \equiv \partial _\mu  A_\nu   - \partial _\nu  A_\mu  $  is the  field strength of the vector field $A_\mu$, which is normally associated with the electromagnetic field, while $P(\phi,X)$ is an arbitrary function of scalar field $\phi$ and its kinetic $X\equiv -\partial^\mu\phi \partial_\mu\phi/2$, which was firstly investigated in the well-known {\it k}-inflation \cite{Armendariz-Picon:1999hyi,Garriga:1999vw}. Various types of $P(\phi,X)$ have been considered in literature, e.g., the canonical type in Ref. \cite{Kanno:2010nr}, the DBI type in Refs. \cite{Do:2011zz,Ohashi:2013pca,Holland:2017cza,Nguyen:2021emx}, the   generalized ghost condensate type in Ref. \cite{Ohashi:2013pca}, the supersymmetric DBI type \cite{Do:2016ofi}, and the {\it k}-inflation type \cite{Do:2020hjf}. Remarkably, the main conclusion obtained in all these models is that the cosmic no-hair conjecture is widely violated.  A main reason for this result is that the existence of gauge-kinetic function $f(\phi)$ will not let the vector field dilutes rapidly during an inflationary phase.

It is worth noting the the non-minimal coupling between the scalar and vector field can be found in various scenarios. One remarkable example worth to be mentioned is that a non-minimal coupling between scalar and vector fields could be responsible for generating  large-scale galactic electromagnetic fields in the present universe since it breaks down the conformal invariance of electromagnetic field as suggested in Refs. \cite{Turner:1987bw,Ratra:1991bn,Bamba:2006ga,Martin:2007ue}. For example, Ratra showed in Ref.  \cite{Ratra:1991bn} that a non-minimal coupling such as $e^{\alpha \phi}F_{\mu \nu } F^{\mu \nu }$ can give raise to the present intergalactic magnetic field that might be as large as ~ $10^{-9} G$. In principle, other non-minimal couplings between the vector and other fields or geometrical quantities can also generate the present large-scale galactic electromagnetic fields due to the fact that they violate the conformal invariance of Maxwell field \cite{Bamba:2006ga}. For example, see Ref. \cite{Bamba:2008ja} for a non-minimal gravitational coupling such as $(1+f(R))F_{\mu \nu } F^{\mu \nu }$, with $f(R)$ is a function of the Ricci scalar $R$. Very interestingly, an anisotropic inflation model based on such non-minimal coupling has been proposed in Ref. \cite{Adak:2016led}. It is worth noting that another anisotropic inflation due to a different non-minimal coupling such as $J^2(X)F_{\mu\nu}F^{\mu\nu}$, with $J(X)$ is a function of kinetic energy of scalar field, has been constructed  in Refs. \cite{Holland:2017cza,Do:2017onf}.

 In harmony with the mentioned models, we would like to propose in this paper one more non-canonical type of $P(\phi,X)$ given by
\begin{equation} \label{Saez-Ballester}
 P(\phi,X)=  w \phi^n  X-V(\phi),
 \end{equation}
 where $w$ and $n$ are free parameters. In this paper, we will consider $w>0$.   Here, the first term was firstly proposed by S\'aez and Ballester  in the so-called S\'aez-Ballester (SB) theory \cite{Saez:1986dil,Saez:1987st}, while the second term is nothing but the potential of $\phi$, which is introduced just for ensuring the existence of inflationary solutions. Various forms of $V(\phi)$ of canonical scalar field have been considered in Ref. \cite{Martin:2013tda}. It is noted that the original version of the SB theory does not contain the potential $V(\phi)$ \cite{Saez:1986dil,Saez:1987st}. Interestingly, some Bianchi metrics have been shown to exist in the SB theory, according to Refs. \cite{Singh:1991vou,Singh}.  In addition, it has been shown in Ref. \cite{Quiros:2022vhm} that the SB theory can be identified with the Einstein-massless-scalar theory. Recently, the SB theory has been discussed extensively, mostly in connection with the late time accelerated expansion issue, e.g., see Refs. \cite{Rasouli:2019axn,Rasouli:2022tjn,Singh:2024zvm} for an incomplete list of relevant literature. Interested readers might want to read an interesting review paper on the leading theoretical approaches to resolve the late time accelerated expansion issue in Ref. \cite{Bamba:2012cp}. Other cosmological aspects of the SB theory can be found in Refs. \cite{Socorro:2009pt,Rasouli:2017glb,Rasouli:2022hnp}.
 
 From a theoretical point of view, ones can argue, according to Refs. \cite{Nojiri:2005pu,Capozziello:2005tf,Elizalde:2008yf}, that the SB theory is nothing but a special case of the standard scalar-tensor theory, whose cosmological implications are very rich, not just for the early universe, but also for the late time one. Very interestingly, by choosing a suitable transformation, ones can be able to reduce the SB theory to the standard scalar-tensor theory  \cite{Nojiri:2005pu,Elizalde:2008yf}. We will come back to this issue later, after deriving all calculations for seeking inflationary solutions of the SB theory as well as for the stability analysis of these solutions.

In harmony with the choice of the SB theory for seeking power-law solutions, the gauge-kinetic function $f(\phi)$ should take the following form,
\begin{equation}
f(\phi) = f_0 \phi^m
\end{equation}
along with that of the potential,
\begin{equation}
V(\phi) =V_0 \phi^ k,
\end{equation}
where $m$ and $k$ are other undetermined constants.  In conclusion, the action of the SB theory non-minimally coupled to the vector field we would like to investigate in this paper is given by
\begin{equation} \label{action}
S = \int {d^4 } x\sqrt {- g} \left[ {\frac{{R}}
{2}  -\frac{1}{2} w \phi^n  \partial_\mu \phi \partial^\mu \phi - V_0 \phi^ k - \frac{1}
{4}f_0^2 \phi ^{2m} F_{\mu \nu } F^{\mu \nu } } \right].
\end{equation}

Our next goal is to figure out anisotropic power-law inflationary solutions to this model. As the first step, the corresponding Einstein field equations can be derived to be
\begin{equation} \label{Einstein}
R_{\mu\nu}-\frac{1}{2}g_{\mu\nu}R - w \phi^n \partial_\mu \phi \partial_\nu \phi + \left(\frac{1}{2} w \phi^n \partial_\rho \phi \partial^\rho \phi +V_0 \phi^ k +\frac{1}{4}f_0^2 \phi^{2m} F^{\rho\sigma}F_{\rho\sigma} \right) g_{\mu\nu} -f_0^2 \phi^{2m} F_{\mu\gamma}F_\nu{}^\gamma =0.
\end{equation}
Then, the corresponding field equation of the scalar field $\phi$ follows with the form,
\begin{equation} \label{scalar}
w \phi^n \square \phi + \frac{1}{2} w n \phi^{n-1}  \partial_\mu \phi \partial^\mu \phi  =  kV_0 \phi^{k-1} +\frac{1}{2}f_0^2 \phi^{2m-1} F^{\mu\nu}F_{\mu\nu}
\end{equation}
along with that of the vector field given by
\begin{equation}\label{vector}
\partial_\mu \left[ \sqrt{-g} f_0^2 \phi^{2m} F^{\mu\nu}\right]=0.
\end{equation}
Here, $\square \equiv \frac{1}{\sqrt{-g}} \partial_\mu \left(\sqrt{-g} \partial^\mu \right)$ is just the d'Alembert operator. 

So far, all related field equations have been addressed accordingly. Now, we must impose a suitable  form of background spacetime for seeking anisotropic power-law solutions. Following the previous studies, e.g., Refs. \cite{Watanabe:2009ct,Kanno:2010nr,Do:2011zz,Ohashi:2013pca,Holland:2017cza,Nguyen:2021emx,Do:2016ofi,Do:2020hjf},  we would like to consider in this paper a homogeneous but anisotropic Bianchi type I spacetime, whose metric is given by
\begin{equation} \label{metric-Bianchi-I}
ds^2 =-dt^2 +\exp\left[ 2\alpha(t) -4\sigma(t) \right] dx^2 +\exp\left[ 2\alpha(t) +2\sigma(t) \right] \left(dy^2+dz^2 \right),
\end{equation}
in harmony with the vector field $A_\mu$, whose configuration is taken as  $A_\mu   = \left( {0,A_x \left( t \right),0,0} \right)$.  In addition, $\phi$ is assumed to be a homogeneous field, i.e., $\phi=\phi(t)$ with $t$ is the cosmic time. In the expression of the Bianchi type I metric shown above, the scale factor $\sigma(t)$ stands for a deviation from the spatial isotropy governed solely by the remaining scale factor $\alpha(t)$. In other words, the absolute value of $\sigma(t)$ should be much smaller than the value of $\alpha(t)$ during an inflationary phase, in order to be consistent with the current observational data of Planck. The existence of spatial hairs is due to the non-vanishing $\sigma(t)$ \cite{Watanabe:2009ct,Kanno:2010nr}. In a case of vanishing $\sigma(t)$, the Bianchi type I metric will reduce to the well-known spatially flat FLRW one. 

As a result,  Eq. (\ref{vector}) can be integrated directly to give a non-trivial solution of vector field such as
\begin{equation} \label{eq5}
\dot A_x\left({t}\right)=p_A f_0^{-2} \phi^{-2m} \exp\left[{-\alpha-4\sigma}\right],
\end{equation}
with $\dot A_x \equiv dA_x/dt$ and $p_A$ is a constant of integration ~\cite{Watanabe:2009ct,Kanno:2010nr}. With the help of this solution, the Einstein field equations can be formulated explicitly as follows 
\begin{align}\label{field-eq-1}
\dot\alpha^2 &=\dot\sigma^2 +\frac{w}{6}  \phi^n \dot\phi^2 +\frac{V_0}{3} \phi^k +\frac{p_A^2 f_0^{-2}}{6} \phi^{-2m} \exp\left[-4\alpha -4\sigma\right] , \\
\label{field-eq-2}
\ddot\alpha  &= -3\dot\alpha^2  +V_0\phi^k  +\frac{p_A^2 f_0^{-2}}{6} \phi^{-2m} \exp\left[-4\alpha -4\sigma\right] ,\\
\label{field-eq-3}
\ddot\sigma &=-3\dot\alpha\dot\sigma +\frac{p_A^2 f_0^{-2}}{3} \phi^{-2m} \exp\left[-4\alpha -4\sigma\right] .
\end{align}
In addition, the corresponding equation of motion of $\phi$ reads
\begin{equation}\label{field-eq-4}
w \phi^n \ddot\phi =-3 w \phi^n \dot\alpha \dot\phi -\frac{w}{2} n \phi^{n-1} \dot\phi^2  -kV_0 \phi^{k-1}+ m p_A^2 f_0^{-2}\phi^{-2m-1} \exp\left[-4\alpha -4\sigma\right] .
\end{equation}
It appears that the evolution and dynamics of the early universe during an inflationary phase can be encoded in the field equations \eqref{field-eq-1}, \eqref{field-eq-2}, \eqref{field-eq-3}, and \eqref{field-eq-4}.
\section{Anisotropic power-law inflationary solutions} \label{sec3}
So far, all field equations have been worked out for the Bianchi type I metric. Now, we would like to figure out analytical power-law solution for the scale factors by taking the following ansatz,
\begin{equation} \label{ansatz}
\alpha(t) = \zeta \log t,\quad \sigma(t) =\eta \log t,\quad \phi(t)= t^l,
\end{equation}
where $\zeta$, $\eta$,  and $l$ are all undetermined constants.  It therefore turns out that
\begin{align}
&\exp\left[ 2\alpha(t) -4\sigma(t) \right] = t^{2\zeta -4\eta} ,\nonumber\\
 & \exp\left[ 2\alpha(t) +2\sigma(t) \right] =t^{2\zeta+2\eta}, \nonumber\\
 & \exp\left[-4\alpha -4\sigma\right] =t^{-4\zeta-4\eta}, \nonumber\\
&\phi^n =  t^{nl}, \quad \phi^{n-1} = t^{(n-1)l},\nonumber\\
& \phi^{-2m} = t^{-2ml},\quad \phi^{-2m-1} = t^{-(2m+1)l},\nonumber\\
 & \phi^k =  t^{kl},\quad  \phi^{k-1} = t^{(k-1)l},
\end{align}
which help us to reduce all differential field equations \eqref{field-eq-1}, \eqref{field-eq-2}, \eqref{field-eq-3}, and \eqref{field-eq-4} to a set of algebraic equations given by
\begin{align}
\zeta^2 t^{-2} =&~\eta^2 t^{-2} +\frac{w}{6} l^2  t^{nl+2l-2} +\frac{V_0}{3} t^{kl}+\frac{p_A^2 f_0^{-2}}{6} t^{-2ml-4\zeta-4\eta} ,\\
-\zeta t^{-2}=& -3\zeta^2 t^{-2}+V_0  t^{kl} + \frac{p_A^2 f_0^{-2}}{6}t^{-2ml-4\zeta-4\eta} ,\\
-\eta t^{-2}= &-3\zeta \eta t^{-2} +\frac{p_A^2 f_0^{-2}}{3} t^{-2ml-4\zeta-4\eta} ,\\
 w  l \left (l-1 \right)t^{nl+l-2} =& -3w l  \zeta  t^{nl+l-2}-\frac{w}{2} n l^2 t^{nl+l-2} -kV_0  t^{kl-l}+mp_A^2 f_0^{-2}  t^{-2ml -l-4\zeta-4\eta}.
\end{align}
Furthermore, under the imposed constraints given by
\begin{align} \label{constraint-1}
nl+2l-2=&-2,\\
\label{constraint-2}
kl=&-2,\\
\label{constraint-3}
-2ml -4\zeta -4\eta =&-2,
\end{align}
the above set of equations can still be reduced to a more simpler set of algebraic equations such as
\begin{align} \label{algebraic-1}
\zeta^2  =&~\eta^2  +\frac{w}{6} l^2    +\frac{u}{3} +\frac{v}{6},\\
\label{algebraic-2}
-\zeta=& -3\zeta^2 +u  + \frac{v}{6},\\
\label{algebraic-3}
-\eta = &-3\zeta \eta  +\frac{v}{3},\\
\label{algebraic-4}
w l\left(l-1 \right)   =& -3w l  \zeta  -\frac{w}{2} n l^2 -k u  +m v,
\end{align}
where we have introduced additional parameters,
\begin{align}
u=&~ V_0,\\
v=&~p_A^2 f_0^{-2} ,
\end{align}
just for convenience. Now, we are going to solve analytically these equations to figure out the corresponding value of $\zeta$ and $\eta$, by which one can judge the visibility of our proposed model. From the first constraint shown in Eq. \eqref{constraint-1}, we have for $l\neq 0$ that
\begin{equation}
n =-2.
\end{equation}
On the other hand, the second constraint \eqref{constraint-2} implies a relation,
\begin{equation}
l =-\frac{2}{k},
\end{equation}
which can help us to reduce the third constraint \eqref{constraint-3} to another relation,
\begin{equation} \label{zeta-eta-relation}
\zeta =-\eta+ \frac{m}{k}  +\frac{1}{2}.
\end{equation}
It turns out that the positivity of $k$ will imply the negativity of $l$, and vice versa, due to the constraint shown in Eq. \eqref{constraint-2}. And, we will only consider $k>0$ in this paper and therefore $l<0$ as a consequence. 
It now becomes clear that the inflationary constraint $\zeta \gg 1$ will require $m \gg k$ since $|\eta|$ has been assumed to be much smaller than $\zeta$ for any viable anisotropic inflationary solutions. As result, Eq. \eqref{algebraic-3} can be solved to give
\begin{equation} \label{definition-of-v}
v = 3 \eta \left(3\zeta-1 \right),
\end{equation}
which if inserted into Eq. \eqref{algebraic-4} will lead to
\begin{equation} \label{definition-of-u}
u= - \frac{\left(3\zeta-1 \right) \left[3m \left( 2k\zeta  -k -2m\right) -4w  \right]}{2k^2}.
\end{equation}
Furthermore, if we plug $v$, $u$, and $\eta$ defined above into either Eq. \eqref{algebraic-1} or \eqref{algebraic-2} we will get the following equation of $\zeta$ such as
\begin{equation}
-6k\zeta \left(k+2m \right) + k^2 +8km +12m^2 +8w  =0,
\end{equation}
here we have ignored trivial solutions $\zeta=0$ and $\zeta=1/3$, which are not consistent with the inflationary phase. Solving this equation will yield a non-trivial solution of $\zeta$ given by
\begin{equation} \label{solution-of-zeta}
\zeta = \frac{k^2 +8km +12m^2 +8w }{6k\left(k+2m \right) },
\end{equation}
by which the corresponding value of $\eta$ will be determined as
\begin{equation}
\eta = -\frac{4w }{3k \left(k+2m \right)}+\frac{1}{3}.
\end{equation}
Now, we will discuss whether the obtained solution is responsible for the inflationary phase. It appears that the positivity of $u$ acquire that
\begin{equation}
6km\zeta -6m^2 -4w <0
\end{equation}
or equivalently
\begin{equation}\label{inequality-w-barphi-1}
w  < \frac{3}{2}m\left(k\zeta-m \right) =- \frac{3}{2}km \left(\eta -\frac{1}{2}\right),
\end{equation}
where the constraint $m\gg k$  and the relation \eqref{zeta-eta-relation} have been used. Hence, if $k$ and $m$ are all positive definite, then $\eta <1/2$ is needed to ensure that the constraint $w >0$ is not violated. On the other hand, the positivity of $v$ will acquire, according to Eq. \eqref{definition-of-v}, that $\eta >0$ because of the assumption that $\zeta \gg1$. Furthermore,  we have approximated values for $\zeta$ and $\eta$, also due to the constraint $m\gg k$, that
\begin{align}
\zeta \simeq &~\frac{m }{k} \gg 1,\\
\eta \simeq & -\frac{2w }{3km}+\frac{1}{3} <\frac{1}{3},
\end{align}
respectively.  Hence, the positivity of $\eta$ will address the following inequality,
\begin{equation} \label{inequality-w-barphi-2}
w <\frac{km}{2}.
\end{equation} 
Cosmologically, any viable anisotropic inflationary solution should have a small spatial anisotropy parameter defined as $\Sigma/H \equiv \dot\sigma /\dot\alpha $  \cite{Watanabe:2009ct,Kanno:2010nr}. For our current solution, it is important to check this criteria. It appears that
\begin{equation}
\frac{\Sigma}{H} \equiv \frac{\dot\sigma}{\dot\alpha} =\frac{\eta}{\zeta} \simeq \eta \frac{ k}{m} < \frac{1}{3}\frac{k}{m} \ll 1,
\end{equation}
as expected. So far, we have successfully  derived the exact anisotropic power-law inflationary solution for the SB theory non-minimally coupled to the vector field. The next important issue we would like to address in the next section is the stability of the obtained solution during the inflationary phase. Before going to the next section, we would like to note an interesting point that the power-law solution obtained in this section can reduce to that derived the KSW model of canonical scalar field with exponential forms of potential and gauge-kinetic function given by \cite{Kanno:2010nr}
\begin{equation}
V(\phi) = V_0 \exp\left[\lambda \phi \right], \quad f(\phi) = f_0 \exp\left[\rho\phi \right],
\end{equation}
where $\lambda$ and $\rho$ are free parameters. Indeed, by taking correspondences such as $k/\sqrt{w} \leftrightarrow \lambda$ and $m/\sqrt{w} \leftrightarrow \rho$, one can easily recover the power-law solution derived in Ref. \cite{Kanno:2010nr}.
However, it is important to note that the setup of fields in Ref. \cite{Kanno:2010nr} is clearly different from that used in the present paper. This indicates the dynamics of fields might not be the same as that of Ref. \cite{Kanno:2010nr}. For example, the scalar field takes a power-law form in the SB theory, while it takes a logarithmic form in the KSW model  \cite{Kanno:2010nr}.  Furthermore, the stability of the SB theory might not be similar to that investigated in Ref. \cite{Kanno:2010nr} due to its non-canonical feature. In addition, the cosmological viability of the SB theory might be different from that of the KSW model. Indeed, it will be shown in the next section that the corresponding dynamical system of the present SB model turns out to be more complicated than that of the SKW model of canonical scalar field \cite{Kanno:2010nr}. Mathematically, it is not always the case that two different dynamical systems would share the same stability property if they admit the same fixed point(s). Detailed analysis needs to be investigated in order to get solid conclusions. We will come back to this issue and give more discussions on it in Sec. \ref{sec5} for completeness.  
\section{Stability analysis: Dynamical system approach} \label{sec4}
\subsection{Anisotropic fixed point}
Following previous works on the stability analysis of anisotropic inflation of the KSW model and its extensions, e.g., see Refs.  \cite{Watanabe:2009ct,Kanno:2010nr,Do:2011zz,Ohashi:2013pca,Holland:2017cza,Do:2016ofi,Do:2020hjf}, we will use the dynamical system approach by introducing the following dimensionless dynamical variables,
\begin{align}
\bar X = \frac{\dot\sigma}{\dot\alpha}, \quad Y =\phi^{\frac{n}{2}} \frac{\dot\phi}{\dot\alpha}, \quad Z =\frac{p_A f^{-1}}{\dot\alpha} \exp[ -2\alpha-2\sigma] ,
\end{align}
along with auxiliary variables defined as follows
\begin{equation}
 U_1= \frac{\lambda}{\lambda+1},\quad U_2= \frac{\rho}{\rho+1},
\end{equation}
where $\lambda$ and $\rho$ are determined, thanks to a hint from Refs. \cite{Do:2021pqk,Bahamonde:2017ize}, as
\begin{equation}
\lambda =\phi^{-\frac{n}{2}}  \frac{\partial_\phi V}{V},\quad \rho =\phi^{-\frac{n}{2}}  \frac{\partial_\phi f}{f}.
\end{equation}
The reason for the introduction of $U_1$ and $U_2$ is due to the non-exponential form of $V(\phi)$ and $f(\phi)$ \cite{Do:2021pqk,Bahamonde:2017ize}. And the reason of the existence of $\phi^{-\frac{n}{2}}$ in the definition of $\lambda$ and $\rho$ is due to the non-minimal coupling between $\phi^n$ and the kinetic term of $\phi$ in the action. This result follows our previous paper \cite{Do:2021pqk}.
Then, we will define the corresponding autonomous equations based on the field equations worked out in the previous section. It is noted that in order to derive the above autonomous equations, we have used the following results,
\begin{equation}
\lambda = \frac{U_1}{1-U_1}=k \phi^{-\frac{n}{2}-1}, \quad \rho =\frac{U_2}{1-U_2}=m \phi^{-\frac{n}{2}-1},\quad \phi^{-1} = \frac{\lambda}{k}  \phi^{\frac{n}{2}} = \frac{\rho}{m}\phi^{\frac{n}{2}} .
\end{equation}
As a result, the corresponding autonomous equations read (see Appendix \ref{appendix1} for detailed derivations)
\begin{align} \label{dyn-eq-1}
\frac{d\bar X}{d\alpha} &= \frac{\ddot\sigma}{\dot\alpha^2} - \bar X \frac{\ddot\alpha}{\dot\alpha^2},\\
\label{dyn-eq-2}
\frac{dY}{d\alpha} &= \frac{n}{2k} \frac{U_1}{1-U_1} Y^2 +\phi^{\frac{n}{2}} \frac{\ddot\phi}{\dot\alpha^2} -Y \frac{\ddot\alpha}{\dot\alpha^2},\\
\label{dyn-eq-3}
\frac{dZ}{d\alpha} &= - Z \left[2\left(\bar X+1\right) +\frac{ m}{k}\frac{U_1}{1-U_1}Y  +\frac{\ddot\alpha}{\dot\alpha^2}  \right],\\
\label{dyn-eq-5}
\frac{d U_1}{d\alpha} &= -\frac{1}{k} \left(\frac{n}{2}+1\right) Y U_1^2,\\
\label{dyn-eq-6}
\frac{d U_2}{d\alpha} & = -\frac{1}{m} \left(\frac{n}{2}+1\right) Y U_2^2.
\end{align}
Here $\alpha \equiv \int \dot\alpha dt$ acts as a dynamical time variable. 
It is noted that the undefined terms $\ddot\alpha/\dot\alpha^2$, $\ddot\sigma/\dot\alpha^2$, and $\phi^{\frac{n}{2}} \ddot\phi/\dot\alpha^2$  in the above autonomous equations can be determined from the field equations of the considered model, i.e., Eqs. \eqref{field-eq-1}, \eqref{field-eq-2}, \eqref{field-eq-3}, and \eqref{field-eq-4}. In particular, we are able to define the following results,
\begin{align}
\frac{\ddot\alpha}{\dot\alpha^2} &= -3\bar X^2 -\frac{w}{2} Y^2 -\frac{Z^2}{3},\\
\frac{\ddot\sigma}{\dot\alpha^2} &= -3\bar X +\frac{Z^2}{3},\\
 \phi^{\frac{n}{2}} \frac{\ddot\phi}{\dot\alpha^2}&= -3Y +\frac{3}{w}  \frac{U_1}{1-U_1} \left(\bar X^2-1 \right)  - \frac{U_1}{2\left(1-U_1\right)} \left(\frac{n}{k}-1\right)Y^2 + \frac{U_1}{w\left (1-U_1\right)} \left( \frac{m}{k}+  \frac{1}{2} \right) Z^2.
\end{align}
Plugging these definitions into Eqs. \eqref{dyn-eq-1}, \eqref{dyn-eq-2}, and \eqref{dyn-eq-3} will yield
\begin{align}\label{dyn-eq-7}
\frac{d\bar X}{d\alpha} =& -3\bar X +\frac{Z^2}{3} + \bar X \left( 3\bar X^2 +\frac{w}{2} Y^2 +\frac{Z^2}{3}\right) ,\\
\label{dyn-eq-8}
\frac{dY}{d\alpha} =& -3Y+ \frac{3}{w}  \frac{U_1}{1-U_1} \left(\bar X^2-1 \right)  + \frac{U_1}{w\left (1-U_1\right)} \left( \frac{m}{k}+  \frac{1}{2} \right) Z^2 \nonumber\\ & +Y \left( \frac{U_1}{1-U_1 } \frac{Y}{2}  + 3\bar X^2 +\frac{w}{2} Y^2 +\frac{Z^2}{3} \right),\\
\label{dyn-eq-9}
\frac{dZ}{d\alpha} =& - Z \left[2\left(\bar X+1\right) +\frac{ m}{k}\frac{U_1}{1-U_1}Y   -3\bar X^2 -\frac{w}{2} Y^2 -\frac{Z^2}{3} \right].
\end{align}
So far, we have worked out the complete dynamical system with the corresponding autonomous equations given by Eqs. \eqref{dyn-eq-5}, \eqref{dyn-eq-6}, \eqref{dyn-eq-7}, \eqref{dyn-eq-8}, and \eqref{dyn-eq-9}. Now, we would like to seek their anisotropic fixed points with $\bar X \neq 0$ from the corresponding equations, 
\begin{equation}
\frac{d\bar X}{d\alpha} =\frac{d Y}{d\alpha} =\frac{dZ}{d\alpha}  =\frac{d U_1}{d\alpha} =\frac{d U_2}{d\alpha} =0.
\end{equation}
It is very straightforward to obtain 
\begin{equation}
n=-2,
\end{equation}
from both equations ${d U_1}/{d\alpha} ={d U_2}/{d\alpha}=0 $. This value of $n$ is indeed identical to that required for the power-law inflation investigated in the previous section. Moreover, it will lead to useful relations, 
\begin{equation} \label{useful-relation}
\lambda = \frac{U_1}{1-U_1}=k, \quad \rho =\frac{U_2}{1-U_2}=m,
\end{equation}
which will be used to derive the below equations. Moreover, these relations imply the corresponding value of $U_1$ and $U_2$ such as
\begin{equation}
U_1 =\frac{k}{k+1} <1,\quad U_2 =\frac{m}{m+1} <1.
\end{equation}
 On the other hand, the equation $dZ/d\alpha =0$ implies that
\begin{equation} \label{dyn-eq-12}
2\left(\bar X+1\right) +m Y   -3\bar X^2 -\frac{w}{2} Y^2 -\frac{Z^2}{3} =0 .
\end{equation}
Another equation $dY/d \alpha =0$ gives
\begin{equation}\label{dyn-eq-14}
-3Y+ \frac{3}{w}  k  \left(\bar X^2-1 \right)  +\frac{1}{w} \left( m+  \frac{k}{2} \right) Z^2 +Y \left(\frac{k}{2}Y +3\bar X^2 +\frac{w}{2}Y^2 +\frac{Z^2}{3} \right)  =0.
\end{equation}
 And the remaining equation $d\bar X/d\alpha =0$ provides
\begin{equation}\label{dyn-eq-15}
-3\bar X + \frac{Z^2}{3}+\bar X \left(3\bar X^2+\frac{w}{2}Y^2 +\frac{Z^2}{3} \right) =0.
\end{equation}
It should be noted that we have ignored all trivial solutions such as $Y=0$, $Z=0$,  $U_1=0$, and $U_2=0$, which would not be consistent with anisotropic fixed points. So far, we have derived three equations, \eqref{dyn-eq-12}, \eqref{dyn-eq-14}, and \eqref{dyn-eq-15}, for three main dynamical variables $\bar X$, $Y$, and $Z^2$. Next step will therefore be solving these equations to figure out anisotropic fixed points. 

 As a result, two Eqs. \eqref{dyn-eq-12} and \eqref{dyn-eq-15} imply a non-trivial relation,
\begin{equation}\label{dyn-eq-16}
Z^2 =3\bar X \left(1-2\bar X -m Y\right),
\end{equation}
by which we are able to find out from Eqs. \eqref{dyn-eq-12} and \eqref{dyn-eq-14} a set of non-trivial solutions of $X$ and $Y$,
\begin{align}
\bar X &= \frac{2\left(k^2+2km -4w\right)}{k^2+8km +12m^2+8w},\\
Y&=- \frac{12\left(k+2m\right)}{k^2+8km +12m^2+8w},
\end{align} 
which is indeed equivalent to the anisotropic power-law solution obtained in Sec. \ref{sec3}. In particular, one can easily verify that $\bar X = \eta/\zeta \equiv \Sigma/H$ and $Y=l/\zeta$. Consequently, the corresponding value of $Z^2$ is defined to be
\begin{equation}
Z^2 = -\frac{18\left(k^2+2km -4w\right) \left(k^2 -4km-12m^2 -8w\right)}{\left(k^2+8km +12m^2+8w\right)^2}.
\end{equation}
The positivity of $Z^2$ will be ensured by the constraints, $m\gg k$ and $w <km/2$, which are required for the existence of the anisotropic inflationary solution as discussed in the previous section.
It is noted that some other fixed points have been ignored since they are not equivalent to the anisotropic power-law solution. In the next section, we will investigate the stability of the above anisotropic fixed point during the inflationary phase.
\subsection{Stability of anisotropic fixed point}
So far, we have successfully constructed the corresponding dynamical system for the SB theory non-minimally coupled to the vector field. Moreover, we have identified the anisotropic fixed point of this dynamical system, which is exactly equivalent to the anisotropic power-law solution derived in the previous section. This result implies that this fixed point and the power-law solution will share the same stability property during the inflationary phase. 

To start the stability investigation, we will perturb the autonomous equations around the anisotropic fixed point as follows
\begin{align}
\frac{d\delta \bar X}{d\alpha} = &-3\delta \bar X +w \bar X Y \delta Y +\frac{2}{3}Z \delta Z,\\
\frac{d\delta Y}{d\alpha} =&~6\bar X \left(\frac{k}{w}+Y\right) \delta \bar X-3\delta Y +\frac{2m}{w}Z \delta Z  -\frac{6}{km\left(1-U_1\right)^2}  \delta U_1,\\
\frac{d\delta Z}{d\alpha} =& -2Z \delta \bar X -mZ \delta Y + \frac{2}{3}Z^2\delta Z -\frac{m}{k}\frac{YZ }{\left(1-U_1\right)^2 }\delta U_1,\\
\frac{d\delta U_1}{d\alpha} =&~ 0,\\
\frac{d\delta U_2}{d\alpha} =&~ 0,
\end{align} 
here we have used the approximated results for the inflationary solutions with $m\gg k \simeq {\cal O}(0.1)$ and $w<km/2$,
\begin{align} \label{appro-1}
\bar X &\simeq \frac{km-2w}{3m^2} <\frac{1}{3}\frac{k}{m} \ll 1,\\
\label{appro-2}
Y&\simeq -\frac{2}{m} <0,\\
\label{appro-3}
Z^2 &\simeq 9\bar X  \ll 1. 
\end{align}
It is clear that $|Y|\ll 1$ for $m\gg 1$.
And a further step is taking exponential perturbations such as
\begin{equation}
\delta \bar X,~\delta Y,~\delta Z,~\delta U_1,~\delta U_2 \sim \exp\left[ \omega \alpha \right],
\end{equation}
which makes the above perturbed equations to the following set of algebraic equation written is a matrix equation,
 \begin{align} \label{stability-equation}
 {\cal M}\left( {\begin{array}{*{20}c}
   A_1  \\
   A_{2}  \\
   A_{3} \\
   A_{4}  \\
   A_{5}  \\
 \end{array} } \right) \equiv \left[ {\begin{array}{*{20}c}
   {-3-\omega} & {w\bar X Y} & {\frac{2}{3}Z } & {0 } & {0}  \\
   { 6\bar X \left(\frac{k}{w}+Y\right) } & {-3-\omega} & {\frac{2m}{w}Z } & { -\frac{6}{km\left(1-U_1\right)^2} } &{0}  \\
     {-2Z  } & {-mZ} & { \frac{2}{3}Z^2 -\omega } & {-\frac{m}{k}\frac{YZ }{\left(1-U_1\right)^2 } } &{0} \\
   {0} & {0} & {0} & { -\omega } &{0} \\
   {0} & {0 } & {0 } &{0}& {-\omega } \\
 \end{array} } \right]  \left( {\begin{array}{*{20}c}
    A_1  \\
   A_{2}  \\
   A_{3} \\
   A_{4}  \\
   A_{5}  \\
 \end{array} } \right) = 0.
\end{align}
Mathematically, this homogeneous set of linear algebraic equations admits non-trivial solutions if and only if
\begin{equation}
\det {\cal M} =0,
\end{equation}
which leads to the corresponding equation of $\omega$ given by
\begin{equation}
\omega^2 f\left(\omega\right) \equiv \omega^2 \left( a_3 \omega^3 +a_2 \omega^2 +a_1 \omega +a_0 \right)=0,
\end{equation}
where
\begin{align}
a_3= &~3w,\\
a_2\simeq &~18w,\\
a_1\simeq &~ 27w +6m^2 Z^2,\\
a_0\simeq &~18m^2Z^2.
\end{align}
Here, only the leading terms in the definition of the coefficients $a_i$ ($i=0-3$) are considered for simplicity.
It is very interesting that all coefficients $a_i$ with $i=0-3$ become positive definite for a positive $w$. This result indicates an important consequence that the corresponding cubic equation, i.e., $f\left(\omega \right)=0$, will only admit non-positive roots of $\omega$, which correspond to a stable anisotropic fixed point. On the other hand, for a negative $w$, we will have $a_3 a_0 <0$, which implies that the corresponding cubic equation $f\left(\omega \right)=0$ will admit at least one positive root $\omega >0$,  resulting that the corresponding anisotropic fixed point is unstable against perturbations. This useful mathematical trick has been used in many of our published papers, e.g., see Refs. \cite{Do:2020hjf,Do:2021pqk}, when we investigated the stability of anisotropic inflationary solutions.

\begin{figure}\centering
	\subfloat[]{\label{1a}\includegraphics[scale=0.64]{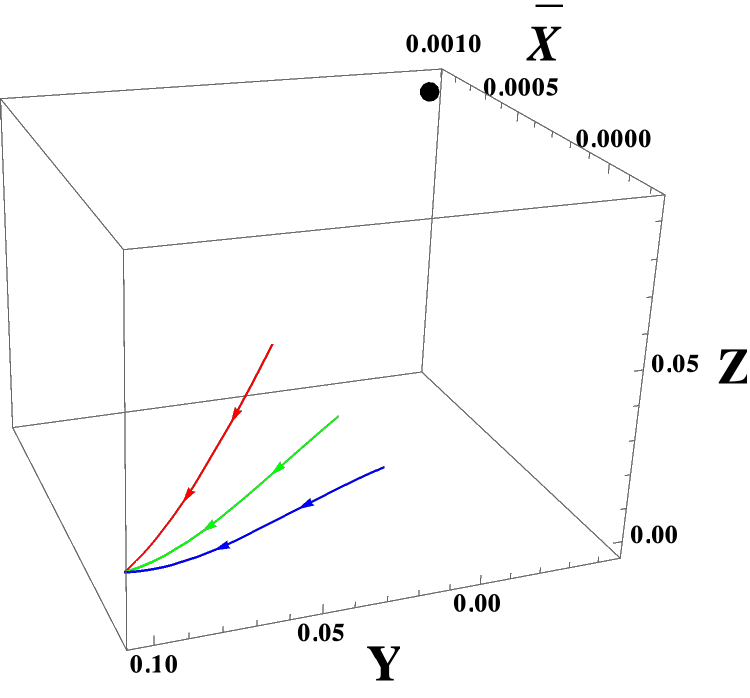}}\hfill
	\subfloat[]{\label{1b}\includegraphics[scale=0.61]{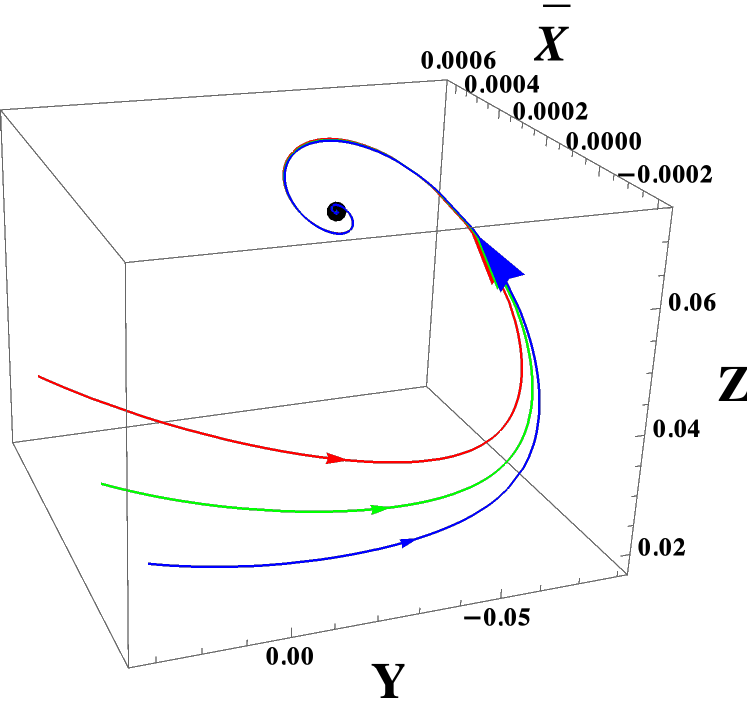}}
	
\caption{\it (Left figure) Unattractive anisotropic fixed point (displayed as a black point) for $k=0.1$, $m=50$, and $w=-1<0$. (Right figure) Attractive anisotropic fixed point (displayed as a black point) for $k=0.1$, $m=50$, and $w=+1>0$. Different colors correspond to different initial conditions of $\bar X$, $Y$, and $Z$.}
\label{fig1}
\end{figure}

To verify the results obtained through the stability analysis, we are going to examine the attractive property of the anisotropic fixed point by doing numerical calculations, similar to the previous studies, e.g., see Refs. \cite{Kanno:2010nr,Do:2011zz,Do:2020hjf,Do:2021pqk}. According to  Fig. \ref{fig1}, it appears that the anisotropic fixed point displayed as a black point with $\left(\bar X,Y,Z\right) \simeq \left(4\times 10^{-4},-4 \times 10^{-2},6\times 10^{-2} \right)$ is indeed an attractive point for $k=0.1$, $m=50$, and $w=+1>0$, due to the fact that three different trajectories starting with different initial conditions all tend to converge toward the black point as the dynamical time variable $\alpha$ evolves. On the other hand, the anisotropic fixed point turns out to be unattractive for $w=-1<0$.  

In short, we have been able to confirmed that the SB theory when non-minimally coupled to the vector field does admit one more counterexample to the cosmic no-hair conjecture. And this result together with our previous papers clearly indicate that the non-trivial coupling $f^2(\phi)F^2$ does play the leading role in breaking down the validity of the cosmic no-hair conjecture, even if the scalar field is of non-canonical forms.  In other words, the spatial anisotropy of the SB theory  non-minimally coupled to the vector field will not be decayed during the inflationary phase. Instead, it would survive until the end of inflationary phase and may therefore leave its imprints on the CMB \cite{Watanabe:2010fh,Dulaney:2010sq,Gumrukcuoglu:2010yc,Watanabe:2010bu,Bartolo:2012sd,Ohashi:2013mka,Ohashi:2013qba,Chen:2014eua,Do:2020ler}.

 \section{Equivalence between the S\'aez-Ballester  and standard scalar-tensor theories} \label{sec5}
 So far, we have presented our own way to investigate the anisotropic power-law inflationary solution along with its stability within the SB theory non-minimally coupled to the vector field. This is basically based on our previous works on non-canonical extensions of the KSW model \cite{Do:2020hjf,Pham:2023evo,Do:2021pqk}.  As noted in the previous section, there has been an effective approach to re-derive all the obtained results through a suitable transformation, by which the SB theory can be shown to be equivalent to the corresponding standard scalar-tensor theory \cite{Nojiri:2005pu,Elizalde:2008yf}. Let us briefly demonstrate in this section about this equivalence. Our analysis basically follows the interesting study published in Refs. \cite{Nojiri:2005pu,Elizalde:2008yf}. 
 
 Let us recall a more general action considered in Refs. \cite{Nojiri:2005pu,Elizalde:2008yf} for a scalar field,
 \begin{equation} \label{general-general-action}
S = \int {d^4 } x\sqrt {- g} \left[ \frac{R}
{2}  -\frac{1}{2}\omega(\phi) \partial_\mu \phi \partial^\mu \phi -V(\phi) \right].
\end{equation}
If one take a field redefinition for $\omega(\phi) >0$ such as
\begin{equation}
\varphi = \int \sqrt{\omega(\phi)} d\phi,
\end{equation}
which is equivalent to 
\begin{equation}
\frac{\partial \varphi }{\partial \phi} =  \sqrt{\omega(\phi)},
\end{equation}
then the above action \eqref{general-general-action} will become as
\begin{equation}
S = \int {d^4 } x\sqrt {- g} \left[ \frac{R}
{2}  -\frac{1}{2} \partial_\mu \varphi \partial^\mu \varphi -V(\varphi) \right],
\end{equation}
which is nothing but the action of standard scalar-tensor theory \cite{Nojiri:2005pu,Elizalde:2008yf}. 

Now, for the specific form of $\omega(\phi) = w \phi^n$ used in the SB theory, we have
\begin{equation}
\varphi  = \int \sqrt{w} \phi^{\frac{n}{2}} d\phi = \frac{2\sqrt{w}}{n+2} \phi^{\frac{n}{2}+1} +\varphi_0,
\end{equation}
 for $n\neq -2$ and 
\begin{equation}
\varphi  = \int \sqrt{w} \phi^{-1} d\phi = \sqrt{w} \log \phi +\varphi_0,
\end{equation}
for $n =  -2$. Here $\varphi_0$ is an integration constant. It becomes clear that the latter case is consistent with the power-law solution found in Sec. \ref{sec3}. In particular, the action \eqref{action} is now rewritten for $n=-2$ as follows
\begin{equation} \label{standard-action}
S = \int {d^4 } x\sqrt {- g} \left[ \frac{R}
{2}  -\frac{1}{2}  \partial_\mu \varphi \partial^\mu \varphi - V_0 e^{\lambda \varphi} - \frac{1}
{4}f_0^2 e^{2\rho \varphi} F_{\mu \nu } F^{\mu \nu }  \right],
\end{equation}
where $\lambda \equiv {k}/{\sqrt{w}}$, $\rho \equiv {m}/{\sqrt{w}}$, $V_0 \to V_0 e^{-\lambda \varphi_0}$, and $f_0 \to f_0 e^{-\rho \varphi_0}$. In addition, the ansatz $\phi(t)= t^l$ now translates to $\varphi(t) = \xi \log t +\varphi_0$ with $\xi \equiv \sqrt{w} l$. Very interestingly, this action is nothing but that of the KSW model \cite{Kanno:2010nr}, consistent with our discussions in Sec. \ref{sec3} that the anisotropic power-law solution obtained in the SB theory is indeed equivalent to that obtained in the KSW model \cite{Kanno:2010nr} via the correspondence such as $k/\sqrt{w} \leftrightarrow \lambda$ and $m/\sqrt{w} \leftrightarrow \rho$,
\begin{equation} \label{solution-of-zeta-KSW}
\zeta = \frac{\lambda^2 +8\lambda \rho  +12\rho^2 +8 }{6\lambda \left(\lambda+2\rho \right) }, \quad \eta = -\frac{4 }{3\lambda \left(\lambda+2\rho \right)}+\frac{1}{3}.
\end{equation}
 It is worth noting that the anisotropic power-law solution of the KSW model has been proven to be stable and attractive during the inflationary phase \cite{Kanno:2010nr}. This result does  confirm our stability analysis presented in the Sec. \ref{sec4}.
\section{Tensor-to-scalar ratio} \label{sec6}
So far, we have successfully constructed a stable and attractive anisotropic inflationary solution to the SB theory in the presence of vector field. One might therefore ask if this solution is viable in the light of the latest data of Planck 2018. Therefore, we will investigate this issue in this section, following our previous works in Refs. \cite{Do:2020hjf,Pham:2023evo}. Indeed, a general expression of tensor-to-scalar ratio for non-canonical anisotropic inflationary models has been derived explicitly in Refs. \cite{Do:2016ofi,Do:2020ler}. Detailed investigations of the tensor-to-scalar ratio within specific models of non-canonical scalar fields such as  the DBI and {\it k}-inflation models can be found in our recent papers \cite{Do:2020hjf,Pham:2023evo}. 

It has been shown in Ref. \cite{Ackerman:2007nb} that once the statistical isotropy of CMB is no longer valid then the scalar power spectrum calculated from the $TT$ correlations will take the following form,
\begin{equation} \label{scalar-power-spectrum-0}
{\cal P}^\zeta_{ k,{\rm ani}} = {\cal P}^{\zeta(0)}_{ k} \left(1+g_\ast \cos^2 \theta \right),
\end{equation}
where ${\cal P}^{\zeta(0)}_{\rm k} $ is nothing but the isotropic scalar power spectrum, which has been worked out for non-canonical scalar field such as \cite{Armendariz-Picon:1999hyi,Garriga:1999vw}
\begin{equation}
{\cal P}^{\zeta(0)}_{ k}  = {\cal P}^{\zeta(0)}_{ k,{\rm nc}} = \frac{1}{8\pi^2 M_p^2} \frac{H^2}{c_s \epsilon},
\end{equation} 
with $\epsilon \equiv -\dot H/H^2$ being the slow-roll parameter, $H$ being the Hubble one, and $c_s$ being the speed of sound of scalar perturbations. On the other hand, according to Ref. \cite{Ackerman:2007nb}, $\theta$ is the angle between the comoving wave number {\bf k} with the privileged direction {\bf V} close to the ecliptic poles, while $g_\ast$ is a constant, whose absolute value is expected to be smaller than one, i.e., $|g_\ast|<1$, and characterizes the deviation from the spatial isotropy. In sort, the term $g_\ast \cos^2 \theta $ plays as a correction due to spatial anisotropies to the scalar power spectrum. Indeed, this correction will disappear for isotropy spaces obeying the cosmological principle with $g_\ast =0$. A number of observational constraints for $g_\ast$ have been figured out via various observations such as that of Planck, WMAP, or  Baryon Oscillation Spectroscopic Survey Data Release 12 \cite{Groeneboom:2009cb,Kim:2013gka,Ramazanov:2013wea,Ramazanov:2016gjl,Sugiyama:2017ggb}. A remarkable point is that these observational constraints all imply that $|g_\ast| <1$. 

From a theoretical point of view, it is important to derive the theoretical expression of $g_\ast$ along with the corresponding power spectra of scalar and tensor perturbations within the context of the KSW model as well as in its non-canonical extensions. It turns out that a systematical investigation for the canonical scalar field has been done in a series of papers by Soda and his colleagues \cite{Watanabe:2010fh,Watanabe:2010bu,Ohashi:2013mka,Ohashi:2013qba}. Related investigations by other people can be found in Refs. \cite{ Dulaney:2010sq,Gumrukcuoglu:2010yc,Bartolo:2012sd,Chen:2014eua}.

For non-canonical scalar fields, the general scalar and tensor power spectra have been derived to be  \cite{Do:2016ofi,Do:2020ler}
\begin{align} \label{scalar-power-spectrum}
{\cal P}^\zeta_{k,\rm nc} &={\cal P}^{\zeta(0)}_{k,\rm nc} \left(1-c_s^5g_\ast^0 \sin^2\theta \right) \simeq {\cal P}^{\zeta(0)}_{k,\rm nc}  \left(1+c_s^5g_\ast^0 \cos^2\theta \right),\\
{\cal P}^h_{k,\rm nc} & ={\cal P}^{h(0)}_{k,\rm nc} \left(1-\frac{\epsilon g_\ast^0}{4} \sin^2\theta \right) \simeq {\cal P}^{h(0)}_{k,\rm nc}  \left(1+\frac{\epsilon g_\ast^0}{4} \cos^2\theta \right),
\end{align}
where ${\cal P}^{h(0)}_{k,\rm nc} $ is the isotropic tensor power spectrum of non-canonical scalar field inflationary models, whose definition is defined to be
\begin{equation}
{\cal P}^{h(0)}_{k,\rm nc} =16 c_s \epsilon {\cal P}^{\zeta(0)}_{k,\rm nc}.
\end{equation} 
Comparing the expression shown in Eq. \eqref{scalar-power-spectrum} with that in Eq. \eqref{scalar-power-spectrum-0} implies that
\begin{equation}
g_\ast = c_s^5 g_\ast^0,
\end{equation}
where $g_\ast^0<0$ is associated with the canonical scalar field, whose explicit definition can be found in Refs. \cite{Watanabe:2010fh,Watanabe:2010bu,Ohashi:2013mka,Ohashi:2013qba}. 
As a consequence, the corresponding full tensor-to-scalar ratio of non-canonical anisotropic inflation reads  \cite{Do:2016ofi,Do:2020ler}
\begin{equation} \label{r-full}
r_{\rm nc} = \frac{{\cal P}^h_{k,\rm nc}}{{\cal P}^\zeta_{k,\rm nc} } = r_{\rm nc}^{\rm iso} \frac{6-\epsilon g_\ast^0}{6-4c_s^5 g_\ast^0},
\end{equation}
where $r_{\rm nc}^{\rm iso}\equiv 16c_s \epsilon $ is the well-known tensor-to-scalar ratio of isotropic inflationary model of non-canonical scalar field \cite{Armendariz-Picon:1999hyi,Garriga:1999vw}. Additionally, the corresponding scalar and tensor spectral indices of non-canonical anisotropic inflation are given by \cite{Do:2016ofi,Do:2020ler}
\begin{align} \label{ns-full}
n_s -1 \simeq &-2\epsilon -\tilde \eta -s +\left( \frac{2}{N_{c_sk}} -5s \right) \frac{2c_s^5 g_\ast^0}{3-2c_s^5 g_\ast^0},\\
n_t \simeq &-2\epsilon,
\end{align}
with $\tilde \eta \equiv \dot\epsilon/(\epsilon H)$, $s \equiv \dot c_s/(c_s H)$, and $N_{c_sk} $ as the e-fold number, which is usually taken to be $60$. Note again that these above formulas have been derived for an arbitrary $P(\phi, X)$.

Given these general results for non-canonical anisotropic inflation with undetermined $c_s$, we now would like to discuss the SB theory with
\begin{equation}
P(\phi,X) =w \phi^n X - V_0 \phi^k.
\end{equation}
Surprisingly, the corresponding $c_s$ of this theory turns out, according to Refs. \cite{Armendariz-Picon:1999hyi,Garriga:1999vw}, to be 
 \begin{equation}
 c_s^2 = \frac{ \partial_X P}{\partial_X \left(2X \partial_X P-P \right)} = 1,
 \end{equation}
 which is clearly identical to that of the canonical scalar field. This is a very special point of the SB theory compared to other models of non-canonical scalar fields such as the DBI and {\it k}-inflation models. And this is also true for a more general extension of SB theory with $P(\phi,X) = h(\phi) X - V(\phi)$, where $h(\phi)$ is an arbitrary function of $\phi$, e.g., see Refs. \cite{Socorro:2009pt,Quiros:2022vhm}. 
 
\begin{figure}\centering
	 \includegraphics[scale=0.9]{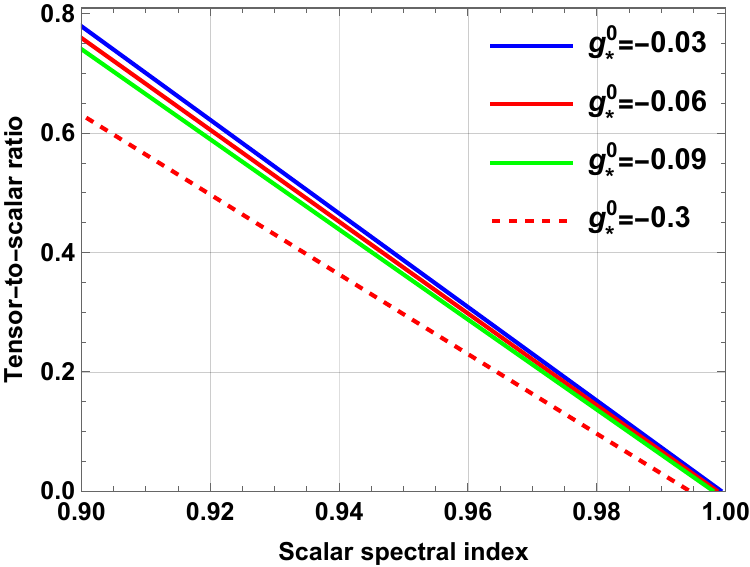}
	
\caption{\it The $n_s - r_{\rm SB} $ diagrams for $g_\ast^0=-0.03$ (blue curve), $g_\ast^0=-0.06$ (red curve), $g_\ast^0=-0.09$ (green curve), and $g_\ast^0 =-0.3$ (dashed red curve) for $4\times 10^{-4}\leq \epsilon \leq 4 \times 10^{-2}$ and $N_{c_sk} =60$, provided that $c_s=1$.}
\label{fig2}
\end{figure}

For the power-law solution found in the previous section, it turns out that
\begin{equation}
\epsilon  = -\frac{\dot H}{H^2} = \frac{1}{\zeta} \simeq \frac{k}{m} \ll 1 \to \tilde \eta = s =0,
\end{equation}
which will imply the corresponding value of the scalar spectral index and tensor-to-scalar ratio given by
\begin{align}
& n_s \simeq 1 -2\epsilon +\frac{4g_\ast^0}{N_{c_sk} \left(3-2g_\ast^0 \right)},\\
&r_{\rm nc} \to r_{\rm SB} = 16 \epsilon \frac{6-\epsilon g_\ast^0}{6-4g_\ast^0},
\end{align}
respectively.  To see if these formulas are consistent with the latest data of the Planck satellite  \cite{Planck:2018vyg,Planck:2018jri}, we will plot the $n_s-r_{\rm SB}$ diagram using the value of parameter $g_\ast^0$ such as $g_\ast^0 =-0.03$, which has been considered in our previous investigations \cite{Do:2020hjf} in order to be  compatible with the recent observational analysis in Refs. \cite{Groeneboom:2009cb,Kim:2013gka,Ramazanov:2013wea,Ramazanov:2016gjl,Sugiyama:2017ggb}. Furthermore, we will also plot this diagram for other values of $g_\ast^0$ such as  $g_\ast^0=-0.06$,  $g_\ast^0=-0.09$, and $g_\ast^0 =-0.3$ just for comparisons. According Fig. \ref{fig2}, we observe the trend that the larger $|g_\ast^0|$ is the smaller $r_{\rm SB}$ and $n_s$ are. However, all analysis made in Refs. \cite{Kim:2013gka,Ramazanov:2013wea,Ramazanov:2016gjl,Sugiyama:2017ggb} put a constraint for $|g_\ast^0|$ that $|g_\ast^0| <0.1$. Only analysis in Ref. \cite{Groeneboom:2009cb} allows $g_\ast^0 \sim - 0.3$. However, even when $g_\ast^0 =-0.3$ is permitted, the corresponding values of $r_{\rm SB}$ and $n_s$ are still higher than the current one observed by the Planck 2018 \cite{Planck:2018vyg,Planck:2018jri}. All these results are due to the result that $c_s=1$. Indeed, it has been shown in our recent papers \cite{Do:2020hjf,Pham:2023evo} that it would be possible to obtain $r < 0.056$ in order to be consistent with the Planck 2018 data \cite{Planck:2018vyg,Planck:2018jri} if the speed of sound, $c_s$, is allowed to be much smaller than one for $g_\ast^0=-0.03$.

\begin{figure}\centering
	  \includegraphics[scale=0.9]{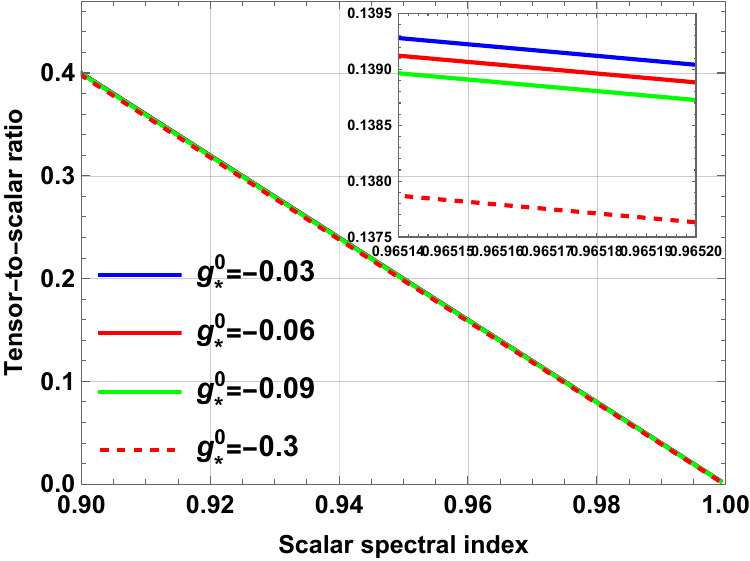}
	
\caption{\it The $ n_s - r_{\rm mSB} $ diagrams for  $ g_\ast^0=-0.03 $ (blue curve), $g_\ast^0=-0.06$ (red curve), $g_\ast^0=-0.09$ (green curve), and $g_\ast^0 =-0.3$ (dashed red curve) for $4\times 10^{-4}\leq \epsilon \leq 4 \times 10^{-2}$ and $N_{c_sk} =60$, provided that $ c_s=0.5 $. It is clear that all different curves become greatly overlapped by the others. (Inset) The enlarged version of the $n_s - r_{\rm mSB} $ diagrams. }
\label{fig3}
\end{figure}

\begin{figure}\centering
	  \includegraphics[scale=0.9]{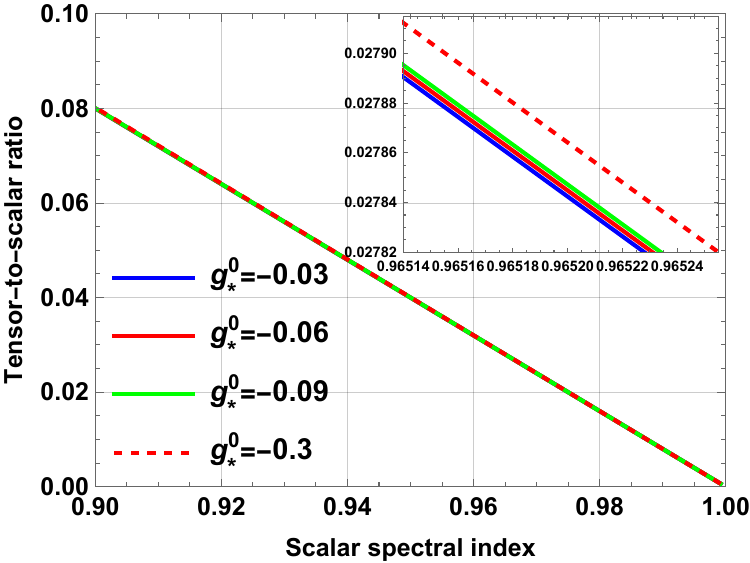}
	
\caption{\it The $n_s - r_{\rm mSB} $ diagrams for $g_\ast^0=-0.03$ (blue curve), $g_\ast^0=-0.06$ (red curve), $g_\ast^0=-0.09$ (green curve), and $g_\ast^0 =-0.3$ (dashed red curve) for $4\times 10^{-4}\leq \epsilon \leq 4 \times 10^{-2}$ and $N_{c_sk} =60$, provided that $c_s=0.1$. It is clear that all different curves become greatly overlapped by the others. (Inset) The enlarged version of the $n_s - r_{\rm mSB} $ diagrams. }
\label{fig4}
\end{figure}

For heuristic reasons, we will plot the $n_s-r_{\rm nc}$ diagram of a modified version of the SB  theory, using two Eqs. \eqref{r-full} and \eqref{ns-full}, for two values  $c_s=0.5$ and $c_s=0.1$, to see if the latest data of the Planck 2018 \cite{Planck:2018vyg,Planck:2018jri}  is met or not. As expected, the value of the tensor-to-scalar ratio will decrease significantly such that the current constraint  $r< 0.056$ by the Planck 2018 data \cite{Planck:2018vyg,Planck:2018jri}  can be easily satisfied with $c_s =0.1$. For the case of $c_s=0.5$, the value of tensor-to-scalar ratio is still higher, so is not suitable. More interestingly,  the value of $g_\ast^0$ does not matter much on the value of the tensor-to-scalar ratio in these cases, compared to $c_s$. All these details can be seen in Figs. \ref{fig4} and \ref{fig3}. Therefore, the SB theory should be modified to have $c_s <1$ in order to be more realistic. For instance, we can propose a simple modified SB (mSB) model such as
\begin{align}
P(\phi,X)=  w \phi^n  X^ {\bar m} - V(\phi),
\end{align}
where $\bar m$ is another constant. One can easily verify that the speed of sound, $c_s$, of this modified version is indeed different from one,
\begin{equation}
c_s^2 = \frac{1}{2 {\bar m}-1}. 
\end{equation}
As a result, the inequality $c_s^2\leq 1$ implies that $\bar m \geq 1$. Additionally, $c_s^2 \ll 1$ as $\bar m \gg 1$. For example, $c_s^2 \sim 10^{-2}$ will acquire that  $2\bar m  \sim 10^{2}$ or equivalently $\bar m \sim 50$. Of course, any negative value of $\bar m$ will lead to $c_s^2 <0$, making the corresponding solution unstable. Complete investigations of this scenario will be our next study and published elsewhere.

\section{Anisotropic power-law inflation for the modified SB model} \label{sec7}
For heuristic reasons, we will investigate in this section the corresponding anisotropic power-law inflationary solution to the modified SB model proposed in the previous section, whose action is given by
\begin{equation} \label{proposed-action}
S = \int {d^4 } x\sqrt {- g} \left[ {\frac{{R}}
{2}  + w \phi^n  X^{\bar m}  - V(\phi) - \frac{1}
{4}f^2 (\phi) F_{\mu \nu } F^{\mu \nu } } \right],
\end{equation}
with $X \equiv -\partial^\mu \phi \partial_\mu \phi/2$ is the kinetic energy of $\phi$. It is clear that the SB theory corresponds to $\bar m =1$. Hence, the KSW model will not be recovered if $\bar m \neq 1$.

As a result, the corresponding Einstein field equation of this model can be defined to be 
\begin{equation}
R_{\mu\nu}-\frac{1}{2}g_{\mu\nu}R - w \bar m \phi^n X^{\bar m-1} \partial_\mu \phi \partial_\nu \phi - \left[ w \phi^n  X^{\bar m}  - V -\frac{1}{4}f^2F_{\rho\sigma}F^{\rho\sigma} \right] g_{\mu\nu} - f^2F_{\mu\gamma}F_\nu{}^\gamma=0,
\end{equation}
along with that of scalar and vector fields given by
\begin{align}
&w \bar m \left[ \phi^n  X^{\bar m-1} \square \phi +\partial_\mu \left( \phi ^n X^{\bar m-1} \right) \partial^\mu \phi \right] = -w n \phi^{n-1}X^{\bar m} + \partial_\phi V + \frac{1}{2}f \partial_\phi f F_{\mu\nu}F^{\mu\nu},\\
&\partial_\mu \left[\sqrt{-g} f^2 F^{\mu\nu} \right] =0,
\end{align}
respectively. Here, $\partial_\phi V  \equiv \partial V /\partial \phi$,  $\partial_\phi f  \equiv \partial f /\partial \phi$, and $\square \phi \equiv \frac{1}{\sqrt{-g}} \partial_\mu \left(\sqrt{-g} \partial ^\mu \phi\right)$.
 Now, using the Bianchi type I metric shown in Eq. \eqref{metric-Bianchi-I} along with the compatible configuration of vector field $A_\mu=(0,A_x(t),0,0)$, we are able to write explicitly the Einstein field equations for $V(\phi)=V_0 \phi^k$ and $f(\phi) =f_0 \phi^m$ as follows
 \begin{align}
 \dot\alpha^2 &= \dot\sigma^2 + \frac{w}{3}\left( 2\bar m -1 \right)\phi^n  \left(\frac{1}{2}\dot\phi^2 \right)^{\bar m} + \frac{V_0 }{3}\phi^k + \frac{p_A^2 f_0^{-2}}{6} \phi^{-2m} \exp[-4\alpha-4\sigma],\\
 \ddot\alpha &=-3\dot\alpha^2 +w \left(\bar m -1 \right) \phi^n  \left(\frac{1}{2}\dot\phi^2 \right)^{\bar m} +V_0 \phi^k +\frac{p_A^2 f_0^{-2}}{6} \phi^{-2m} \exp[-4\alpha-4\sigma],\\
 \ddot\sigma &= -3\dot\alpha\dot\sigma +\frac{p_A^2 f_0^{-2}}{3} \phi^{-2m} \exp[-4\alpha-4\sigma].
 \end{align}
 Additionally, the corresponding equation of motion of scalar field turns out to be
 \begin{align}
 w \bar m \phi^n \left(\frac{1}{2} \dot\phi ^2 \right)^{\bar m-1} \left[\left(2\bar m-1 \right) \ddot\phi +3\dot\alpha \dot\phi  \right] = &- wn \left(2\bar m-1 \right) \phi^{n-1} \left(\frac{1}{2} \dot\phi ^2 \right)^{\bar m} -kV_0 \phi^{k-1} \nonumber\\
 & +m p_A^2 f_0^{-2}  \phi^{-2m-1} \exp[-4\alpha-4\sigma].
 \end{align}
 It should be noted that we have used the solution of vector field shown in Eq. \eqref{eq5}.  It appears that all these equations will recover that of the SB theory derived in Sec. \ref{sec3} once the limit $\bar m \to 1$ is taken.
 
 Next, we will seek the corresponding anisotropic power-law solution to this model by using the same ansatz \eqref{ansatz}, i.e.,
 \begin{equation}
\alpha(t) = \zeta \log t,\quad \sigma(t) =\eta \log t,\quad \phi(t)= t^l.
\end{equation}
As a result, the corresponding set of algebraic equations is given by
\begin{align} \label{mSB-1}
\zeta^2&=\eta^2 +\frac{w}{3} \left(2\bar m -1\right) \left(\frac{l^2}{2} \right)^{\bar m} +\frac{u}{3}+\frac{v}{6},\\
\label{mSB-2}
-\zeta &= -3\zeta^2 +w \left(\bar m -1 \right) \left(\frac{l^2}{2} \right)^{\bar m} +u+\frac{v}{6},\\
\label{mSB-3}
-\eta&= -3\zeta \eta +\frac{v}{3},\\
\label{mSB-4}
w\bar m  \left(\frac{l^2}{2} \right)^{\bar m -1 } \left[\left(2\bar m-1 \right) l \left(l-1\right) +3\zeta l \right]& =-wn \left(2\bar m-1 \right) \left(\frac{l^2}{2} \right)^{\bar m } -k u +m v,
\end{align}
provided the associated constraint equations,
\begin{align}
\left( n+2\bar m \right)l -2\bar m &=-2,\\
kl&=-2,\\
ml +2\zeta +2\eta &=1.
\end{align}
Note that the definition of $u$ and $v$ can be found in Sec. \ref{sec3}. From the constraint equations, it appears that
\begin{align}
l &=-\frac{2}{k},\\
\eta & = -\zeta +\frac{m}{k} +\frac{1}{2},\\
n&= -\left(k+2 \right) \bar m +k.
\end{align}
Hence, similar to the case of SB theory $m \gg k~ (>0)$ is still the constraint for the existence of anisotropic inflationary solutions of the modified SB model having a small spatial anisotropy $|\eta| \ll \zeta$. For convenience, we will introduce an additional parameter $\gamma$ such as
\begin{equation}
\gamma \equiv  \left(\frac{l^2}{2} \right)^{\bar m} = \left(\frac{2}{k^2} \right)^{\bar m}.
\end{equation}
Thanks to the relations,
\begin{align}
v&= 3\eta \left(3\zeta-1\right),\\
u&= -\frac{1}{2k^2} \left[ 18km \zeta^2 - \left(18m^2 +6w \bar m\gamma k^2+15km \right)\zeta + 6m^2 +2w \gamma  \left(2\bar m-1 \right) k^2 +3km \right],
\end{align}
which are found from Eqs. \eqref{mSB-3} and \eqref{mSB-4}, respectively, we are able to reduce both Eqs. \eqref{mSB-1} and \eqref{mSB-2} to an equation of $\zeta$ such as
\begin{equation}
6k\left(k+2m\right)\zeta - k^2 -8km -12m^2 -4w \bar m \gamma k^2=0.
\end{equation}
Solving this equation gives us a non-trivial solution of $\zeta$,
\begin{equation}
\zeta = \frac{ k^2 +8km +12m^2 + 4w \bar m \gamma k^2}{6k\left(k+2m\right)},
\end{equation}
which will be further reduced to 
\begin{equation}
\zeta = \frac{ k^2 +8km +12m^2 +4 w \bar m k^2\left( \frac{\sqrt{2}}{k} \right)^{2 \bar m}} {6k\left(k+2m\right)}.
\end{equation}
This solution will definitely reduce to that shown in Eq. \eqref{solution-of-zeta} in the limit $\bar m \to 1$. 
 It is clear that $k^{-2\bar m+2} \gg 1$ if $\bar m \gg 1$ and $0<k \ll 1$. Consequently, we have
 \begin{equation} \label{solution-of-zeta-modified}
 \zeta \simeq \frac{m}{k} + \frac{w \bar m k }{3m} \left( \frac{\sqrt{2}}{k} \right)^{2 \bar m}.
 \end{equation}
 Hence, we observe, by comparing with $\zeta \simeq m/k$ found in the SB theory in Sec. \ref{sec3},  that the value of $\zeta$ is significantly  enhanced by the second term in the right hand side of Eq. \eqref{solution-of-zeta-modified}. However, we still want to keep $\eta $ small in order to be compatible with observations, then the correction term of $\zeta$ should not be arbitrarily large. This indicates that the usual choice $0<k\ll 1$ (or equivalently $0<\lambda\ll 1$) as selected in the previous section as well as in the original paper \cite{Kanno:2010nr}, is indeed inconsistent with the smallness requirement of $\eta$. It appears that another choice, $k > \sqrt{2}$, seems to be the only suitable one for this modified SB model.  Indeed, it turns out that the inequality, $\sqrt{2}/k <1$, implies that
 \begin{equation}
 \left( \frac{\sqrt{2}}{k} \right)^{2\bar m} \ll 1 \quad {\text {for}} ~ \bar m \gg 1.
 \end{equation}
 Hence, we will have for this choice the following approximated values of the anisotropic power-law inflationary solution,
 \begin{equation}
 \zeta \simeq \frac{m}{k} \gg 1, \quad \eta \simeq \frac{1}{3} \ll \zeta, \quad \frac{\Sigma}{H} \simeq \frac{1}{3}\frac{k}{m} \ll 1.
 \end{equation}
 
  So far, we have successfully demonstrated that the simple modified SB model \eqref{proposed-action} can admit an anisotropic power-law inflationary solution having a small spatial anisotropy. Detailed investigation of its stability will be our future study and will be published elsewhere. 
\section{Conclusions} \label{final}
We have investigated whether a Bianchi type I power-law inflationary solution exists in the SB  theory \cite{Saez:1986dil,Saez:1987st,Quiros:2022vhm,Rasouli:2022hnp,Rasouli:2019axn,Rasouli:2022tjn,Singh:2024zvm} non-minimally coupled to a vector field or not. As a result, it has been shown that such a solution does appear in this scenario for a suitable setup of fields. By transforming the field equations into the corresponding dynamical system, we have been able to confirm the obtained solution is indeed stable against field perturbations. Furthermore, it has been shown to be an attractor fixed point, meaning that a state of universe at the end of inflationary phase would remain spatially anisotropic, regardless of initial conditions. All these results indicate that the Hawking cosmic no-hair conjecture is no longer valid in this scenario, which  seems to be nothing but a new non-canonical extension of the KSW model \cite{Watanabe:2009ct,Kanno:2010nr}, along with the others already found in the previous papers \cite{Do:2011zz,Ohashi:2013pca,Holland:2017cza,Nguyen:2021emx,Do:2016ofi,Do:2020hjf,Pham:2023evo}. Very interestingly, the considered S\'aez-Ballester theory has been shown to be equivalent to the standard scalar-vector theory via a suitable field redefinition. This means that results obtained in this paper can be regarded a non-trivial reconfirmation of that derived in the KSW model  \cite{Kanno:2010nr}.

To see if the obtained solution is viable or not in the light of the Planck 2018 data, we have considered its corresponding tensor-to-scalar ratio. Due to the interesting result that the speed of sound $c_s$ of the SB theory turns out to be one, which is identical to that of canonical scalar field model \cite{Watanabe:2009ct,Kanno:2010nr}, the value of this ratio is higher than expected. This result clearly implies that the role of $c_s$ cannot be ignored in the context of KSW anisotropic inflation. In particular, $c_s<1$ would reduce the value of the tensor-to-scalar ratio to proper values consistent with the Planck 2018 data as pointed out in our previous papers \cite{Do:2020hjf,Pham:2023evo}. To be more specific with our expectation, we have plotted this ratio for two specific values of $c_s$ such as  $c_s =0.5$ and $c_s=0.1$. According to the obtained numerical results, $c_s \sim 0.1$ seems to be a desired value. For heuristic reasons, we have proposed a simple modification of the SB theory, whose $c_s$ is no longer one.  Its anisotropic power-law inflation has been shown to exist as expected. Detailed investigations of the stability of this modified SB model will be our next study and will be published elsewhere. Finally, we hope that our current research would be useful to further works on anisotropic inflation.
\begin{acknowledgments}
We would like to thank a referee very much for his/her critical and constructive comments. One of us (T.Q.D.) would like to thank Prof. W. F. Kao very much for his useful comments and his previous collaborations on anisotropic inflation. Tuyen M. Pham is appreciated very much for his useful assistance.  This study is funded by the Vietnam National Foundation for Science and Technology Development (NAFOSTED) under grant number 103.01-2023.50 (T.Q.D., P.V.D., and D.H.N.).  
\end{acknowledgments}
\appendix
\section{Autonomous equations} \label{appendix1}
Autonomous equations are defined as follows
\begin{align} \label{App-dyn-eq-1}
\frac{d\bar X}{d\alpha} = \frac{d \bar X}{dt} \frac{dt}{d\alpha} = \dot\alpha^{-1} \frac{d}{dt} \left(\frac{\dot\sigma}{\dot\alpha} \right) = \frac{\ddot\sigma}{\dot\alpha^2} - \bar X \frac{\ddot\alpha}{\dot\alpha^2},
\end{align}
\begin{align}\label{App-dyn-eq-2}
\frac{dY}{d\alpha} = \frac{dY}{dt}\frac{dt}{d\alpha} &= \dot\alpha^{-1} \frac{d}{dt} \left(\frac{\phi^{\frac{n}{2}} \dot\phi}{\dot\alpha} \right) \nonumber\\
&= \frac{n}{2} \phi^{- \frac{n}{2}-1} Y^2 +\phi^{\frac{n}{2}} \frac{\ddot\phi}{\dot\alpha^2} -Y \frac{\ddot\alpha}{\dot\alpha^2} \nonumber\\
&= \frac{n}{2} \frac{\lambda}{k} Y^2 +\phi^{\frac{n}{2}} \frac{\ddot\phi}{\dot\alpha^2} -Y \frac{\ddot\alpha}{\dot\alpha^2} \nonumber\\
&= \frac{n}{2k} \frac{U_1}{1-U_1} Y^2 +\phi^{\frac{n}{2}} \frac{\ddot\phi}{\dot\alpha^2} -Y \frac{\ddot\alpha}{\dot\alpha^2},
\end{align}
\begin{align}\label{App-dyn-eq-3}
\frac{dZ}{d\alpha} = \frac{dZ}{dt}\frac{dt}{d\alpha} &= \dot\alpha^{-1} \frac{d}{dt} \left\{\frac{p_A f^{-1}}{\dot\alpha} \exp[ -2\alpha-2\sigma] \right\} 
\nonumber\\
&= - Z \left[2\left(\bar X+1\right) +m\phi^{-1}\frac{\dot\phi}{\dot\alpha}  +\frac{\ddot\alpha}{\dot\alpha^2}  \right] \nonumber\\
&= - Z \left[2\left(\bar X+1\right) +\frac{m}{k}\lambda Y  +\frac{\ddot\alpha}{\dot\alpha^2}  \right] \nonumber\\
&= - Z \left[2\left(\bar X+1\right) +\frac{ m}{k}\frac{U_1}{1-U_1}Y  +\frac{\ddot\alpha}{\dot\alpha^2}  \right],
\end{align}
\begin{align}\label{App-dyn-eq-5}
\frac{d U_1}{d\alpha} = \frac{dU_1}{dt}\frac{dt}{d\alpha} &=  \dot\alpha^{-1}  \frac{d}{dt} \left(\frac{\lambda}{\lambda+1} \right) \nonumber\\
&= -k\left(\frac{n}{2}+1\right) \phi^{-\frac{n}{2}-2} \frac{\dot\phi}{\dot\alpha} \frac{1}{(\lambda+1)^2} \nonumber\\
&= -k\left(\frac{n}{2}+1\right) \frac{\lambda^2}{k^2} Y \left(1-U_1\right)^2 \nonumber\\
&= -\frac{1}{k} \left(\frac{n}{2}+1\right) Y U_1^2,
\end{align}
\begin{align}\label{App-dyn-eq-6}
\frac{d U_2}{d\alpha} = \frac{dU_2}{dt}\frac{dt}{d\alpha} &= \dot\alpha^{-1} \frac{d}{dt} \left(\frac{\rho}{\rho+1} \right) \nonumber\\
&= -m\left(\frac{n}{2}+1\right) \phi^{-\frac{n}{2}-2} \frac{\dot\phi}{\dot\alpha} \frac{1}{(\rho+1)^2} \nonumber\\
&= -m\left(\frac{n}{2}+1\right) \frac{\rho^2}{m^2} Y \left(1-U_2\right)^2 \nonumber\\
&= -\frac{1}{m} \left(\frac{n}{2}+1\right) Y U_2^2.
\end{align}


\begin{thebibliography}{99} 
				
		\bibitem{Schwarz:2015cma}
		D.~J.~Schwarz, C.~J.~Copi, D.~Huterer, and G.~D.~Starkman,
		CMB anomalies after Planck,
		Class. Quant. Grav. {\bf 33}, 184001 (2016)
		[arXiv:1510.07929].
		
		\bibitem{Buchert:2015wwr}
		T.~Buchert, A.~A.~Coley, H.~Kleinert, B.~F.~Roukema, and D.~L.~Wiltshire,
		Observational challenges for the standard FLRW model,
		Int. J. Mod. Phys. D \textbf{25}, 1630007 (2016)
		[arXiv:1512.03313].
		
\bibitem{Perivolaropoulos:2021jda}
L.~Perivolaropoulos and F.~Skara,
Challenges for \ensuremath{\Lambda}CDM: An update,
New Astron. Rev. \textbf{95}, 101659 (2022)
[arXiv:2105.05208].

		
\bibitem{Abdalla:2022yfr}
E.~Abdalla, G.~Franco Abell\'an, A.~Aboubrahim, A.~Agnello, O.~Akarsu, Y.~Akrami, G.~Alestas, D.~Aloni, L.~Amendola and L.~A.~Anchordoqui, \textit{et al.}
Cosmology intertwined: A review of the particle physics, astrophysics, and cosmology associated with the cosmological tensions and anomalies,
JHEAp \textbf{34}, 49 (2022)
[arXiv:2203.06142].
		
\bibitem{Aluri:2022hzs}
P.~K.~Aluri, P.~Cea, P.~Chingangbam, M.~C.~Chu, R.~G.~Clowes, D.~Hutsem\'ekers, J.~P.~Kochappan, A.~M.~Lopez, L.~Liu, and N.~C.~M.~Martens, \textit{et al.},
Is the observable Universe consistent with the cosmological principle?,
Class. Quant. Grav. \textbf{40},  094001 (2023)
[arXiv:2207.05765].

		\bibitem{WMAP:2012nax}
		G.~Hinshaw \textit{et al.},
		Nine-Year Wilkinson Microwave Anisotropy Probe (WMAP) Observations: Cosmological Parameter Results,
		Astrophys. J. Suppl. \textbf{208}, 19 (2013)	
		[arXiv:1212.5226].

\bibitem{Planck:2019evm}
Y.~Akrami \textit{et al.} [Planck],
Planck 2018 results. VII. Isotropy and Statistics of the CMB,
Astron. Astrophys. \textbf{641}, A7 (2020)
[arXiv:1906.02552].

\bibitem{Kester:2023qmm}
C.~E.~Kester, A.~Bernui, and W.~S.~Hip\'olito-Ricaldi,
Probing the statistical isotropy of the universe with Planck data of the cosmic microwave background,
Astron. Astrophys. \textbf{683}, A176 (2024)
[arXiv:2310.02928].


\bibitem{Jones:2023ncn}
J.~Jones, C.~J.~Copi, G.~D.~Starkman, and Y.~Akrami,
The Universe is not statistically isotropic,
arXiv:2310.12859.

\bibitem{Krishnan:2021dyb}
C.~Krishnan, R.~Mohayaee, E.~\'O.~Colg\'ain, M.~M.~Sheikh-Jabbari, and L.~Yin,
Does Hubble tension signal a breakdown in FLRW cosmology?,
Class. Quant. Grav. \textbf{38},  184001 (2021)
[arXiv:2105.09790].

\bibitem{Krishnan:2021jmh}
C.~Krishnan, R.~Mohayaee, E.~\'O.~Colg\'ain, M.~M.~Sheikh-Jabbari, and L.~Yin,
Hints of FLRW breakdown from supernovae,
Phys. Rev. D \textbf{105},  063514 (2022)
doi:10.1103/PhysRevD.105.063514
[arXiv:2106.02532].


\bibitem{Colin:2019opb}
J.~Colin, R.~Mohayaee, M.~Rameez, and S.~Sarkar,
Evidence for anisotropy of cosmic acceleration,
Astron. Astrophys. \textbf{631}, L13 (2019)
[arXiv:1808.04597].

\bibitem{Secrest:2020has}
N.~J.~Secrest, S.~von Hausegger, M.~Rameez, R.~Mohayaee, S.~Sarkar, and J.~Colin,
A Test of the Cosmological Principle with Quasars,
Astrophys. J. Lett. \textbf{908}, L51 (2021)
[arXiv:2009.14826].

\bibitem{Rameez:2024xsn}
M.~Rameez,
Anisotropy in the cosmic acceleration inferred from supernovae,
arXiv:2411.14758.

\bibitem{Boubel:2024cmh}
P.~Boubel, M.~Colless, K.~Said, and L.~Staveley-Smith,
Testing anisotropic Hubble expansion,
arXiv:2412.14607.


		\bibitem{Gibbons:1977mu}
		G.~W.~Gibbons and S.~W.~Hawking,
		Cosmological event horizons, thermodynamics, and particle creation,
		Phys. Rev. D \textbf{15}, 2738-2751 (1977).
		
		\bibitem{Hawking:1981fz}
		S.~W.~Hawking and I.~G.~Moss,
		Supercooled phase transitions in the very early universe,
		Phys. Lett. B \textbf{110}, 35 (1982).
	
  \bibitem{Wald:1983ky} 
  R.~M.~Wald,
  Asymptotic behavior of homogeneous cosmological models in the presence of a positive cosmological constant,
 Phys. Rev. D {\bf 28}, 2118 (1983).
 

\bibitem{Barrow:1987ia} 
  J.~D.~Barrow,
  Cosmic no hair theorems and inflation,
  Phys. Lett. B {\bf 187}, 12 (1987).
 
  \bibitem{Mijic:1987bq}
M.~Mijic and J.~A.~Stein-Schabes,
A no-hair theorem for $R^{2}$ models,
Phys. Lett. B \textbf{203}, 353 (1988).

\bibitem{Kitada:1991ih} 
  Y.~Kitada and K.~i.~Maeda,
  Cosmic no hair theorem in power law inflation,
  Phys. Rev.  D {\bf 45}, 1416 (1992).

 
\bibitem{Maleknejad:2012as}
A.~Maleknejad and M.~M.~Sheikh-Jabbari,
Revisiting cosmic no-hair theorem for inflationary settings,
Phys. Rev. D \textbf{85}, 123508 (2012)
[arXiv:1203.0219].

\bibitem{Kleban:2016sqm} 
  M.~Kleban and L.~Senatore,
  Inhomogeneous anisotropic cosmology,
  J. Cosmol. Astropart. Phys. {\bf 10} (2016) 022 [arXiv:1602.03520]. 
  
\bibitem{East:2015ggf} 
  W.~E.~East, M.~Kleban, A.~Linde, and L.~Senatore,
  Beginning inflation in an inhomogeneous universe,
  J. Cosmol. Astropart. Phys.  {\bf 09} (2016) 010 [arXiv:1511.05143].

\bibitem{Carroll:2017kjo} 
  S.~M.~Carroll and A.~Chatwin-Davies,
  Cosmic equilibration: A holographic no-hair theorem from the generalized second law,
  Phys.  Rev. D {\bf 97},  046012 (2018)
  [arXiv:1703.09241].
  
  
		\bibitem{Kaloper:1991rw}
		N.~Kaloper,
		Lorentz Chern-Simons terms in Bianchi cosmologies and the cosmic no hair conjecture,
		Phys. Rev. D \textbf{44}, 2380 (1991).


\bibitem{Barrow:2005qv} 
  J.~D.~Barrow and S.~Hervik,
  Anisotropically inflating universes,
  Phys.\ Rev.\ D {\bf 73}, 023007 (2006)
  [gr-qc/0511127].
		
		  
		\bibitem{Watanabe:2009ct}
		M.~a.~Watanabe, S.~Kanno, and J.~Soda,
		Inflationary universe with anisotropic hair,
		Phys. Rev. Lett. \textbf{102}, 191302 (2009)
		[arXiv:0902.2833].
		
		\bibitem{Kanno:2010nr}
		S.~Kanno, J.~Soda, and M.~a.~Watanabe,
		Anisotropic power-law inflation,
		J. Cosmol. Astropart. Phys. \textbf{12}, 024 (2010)
		[arXiv:1010.5307].
				
					
		\bibitem{Tahara:2018orv}
		H.~W.~H.~Tahara, S.~Nishi, T.~Kobayashi, and J.~Yokoyama,
		Self-anisotropizing inflationary universe in Horndeski theory and beyond,
		J. Cosmol. Astropart. Phys. \textbf{07}, 058 (2018)
		[arXiv:1805.00186].
		
		\bibitem{Starobinsky:2019xdp}
		A.~A.~Starobinsky, S.~V.~Sushkov, and M.~S.~Volkov,
		Anisotropy screening in Horndeski cosmologies,
		Phys. Rev. D \textbf{101}, 064039 (2020)
		[arXiv:1912.12320].
		
		\bibitem{Galeev:2021xit}
		R.~Galeev, R.~Muharlyamov, A.~A.~Starobinsky, S.~V.~Sushkov, and M.~S.~Volkov,
		Anisotropic cosmological models in Horndeski gravity,
		Phys. Rev. D \textbf{103}, 104015 (2021)
		[arXiv:2102.10981].
		
\bibitem{Nojiri:2022idp}
S.~Nojiri, S.~D.~Odintsov, V.~K.~Oikonomou, and A.~Constantini,
Formalizing anisotropic inflation in modified gravity,
Nucl. Phys. B \textbf{985}, 116011 (2022)
[arXiv:2210.16383].


		  
\bibitem{Starobinsky:1982mr}
A.~A.~Starobinsky, Isotropization of arbitrary cosmological expansion given an effective cosmological constant, JETP Lett. \textbf{37}, 66 (1983).


\bibitem{Muller:1989rp}
 V.~Muller, H.~J.~Schmidt, and A.~A.~Starobinsky, Power law inflation as an attractor solution for inhomogeneous cosmological models,
Class. Quant. Grav. \textbf{7}, 1163 (1990).

\bibitem{Barrow:1984zz}
J.~D.~Barrow and J.~Stein-Schabes,
Inhomogeneous cosmologies with cosmological constant,
Phys. Lett. A \textbf{103}, 315 (1984).


\bibitem{Jensen:1986nf}
 L.~G.~Jensen and J.~A.~Stein-Schabes,
Is inflation natural?,
Phys. Rev. D \textbf{35}, 1146 (1987).

\bibitem{SteinSchabes:1986sy}
J.~A.~Stein-Schabes,
Inflation in spherically symmetric inhomogeneous models,
Phys. Rev. D \textbf{35}, 2345 (1987).


 \bibitem{Ellis:1968vb} 
  G.~F.~R.~Ellis and M.~A.~H.~MacCallum,
  A Class of homogeneous cosmological models,
  Commun.\ Math.\ Phys.\  {\bf 12}, 108 (1969).
  
  
		\bibitem{Pitrou:2008gk}
		C.~Pitrou, T.~S.~Pereira, and J.~P.~Uzan,
		Predictions from an anisotropic inflationary era,
		J. Cosmol. Astropart. Phys. \textbf{04}, 004 (2008)
		[arXiv:0801.3596].
		
		\bibitem{Gumrukcuoglu:2007bx}
		A.~E.~Gumrukcuoglu, C.~R.~Contaldi, and M.~Peloso,
		Inflationary perturbations in anisotropic backgrounds and their imprint on the CMB,
		J. Cosmol. Astropart. Phys. \textbf{11}, 005 (2007)
		[arXiv:0707.4179].



		\bibitem{Starobinsky:1980te}
		A.~A.~Starobinsky,
		A new type of isotropic cosmological models without singularity,
		Phys. Lett. B \textbf{91}, 99 (1980).
		
		
		\bibitem{Guth:1980zm}
		A.~H.~Guth,
		The inflationary universe: A possible solution to the horizon and flatness problems,
		Phys. Rev. D \textbf{23}, 347 (1981).
		
		\bibitem{Linde:1981mu}
		A.~D.~Linde,
		A new inflationary universe scenario: A possible solution of the horizon, flatness, homogeneity, isotropy and primordial monopole problems,
		Phys. Lett. B \textbf{108}, 389 (1982).
		
		\bibitem{Lyth:2009zz}
		D.~H.~Lyth and A.~R.~Liddle,
		The primordial density perturbation: Cosmology, inflation and the origin of structure, Cambridge University Press (2009).
		
		\bibitem{Planck:2018vyg}
		N.~Aghanim \textit{et al.}, 
		Planck 2018 results. VI. Cosmological parameters,
		Astron. Astrophys. \textbf{641}, A6 (2020)
		[arXiv:1807.06209].
		
		\bibitem{Planck:2018jri}
		Y.~Akrami \textit{et al.},
		Planck 2018 results. X. Constraints on inflation,
		Astron. Astrophys. \textbf{641}, A10 (2020)
		[arXiv:1807.06211].
		

			
\bibitem{Odintsov:2023weg}
S.~D.~Odintsov, V.~K.~Oikonomou, I.~Giannakoudi, F.~P.~Fronimos, and E.~C.~Lymperiadou,
Recent advances on inflation,
Symmetry \textbf{15}, 1701 (2023)
[arXiv:2307.16308].

		
\bibitem{Martin:2013tda}
J.~Martin, C.~Ringeval, and V.~Vennin,
Encyclop\ae{}dia Inflationaris: Opiparous Edition,
Phys.Dark Univ. {\bf 46}, 101653 (2024);
[arXiv:1303.3787].
  
 

		\bibitem{Kao:2009zzb}
		W.~F.~Kao and I.~C.~Lin,
		Anisotropically inflating universes in a scalar-tensor theory,
		Phys. Rev. D \textbf{79}, 043001 (2009).
		
		\bibitem{Chang:2011zzb}
		C.~Chang, W.~F.~Kao, and I.~C.~Lin,
		Stability analysis of the Lorentz Chern-Simons expanding solutions,
		Phys. Rev. D \textbf{84}, 063014 (2011).
		
	
		
\bibitem{Watanabe:2010fh}
M.~a.~Watanabe, S.~Kanno, and J.~Soda,
The nature of primordial fluctuations from anisotropic inflation,
Prog. Theor. Phys. \textbf{123}, 1041 (2010)
[arXiv:1003.0056].
 
\bibitem{Dulaney:2010sq} 
  T.~R.~Dulaney and M.~I.~Gresham,
 Primordial power spectra from anisotropic inflation,
  Phys.\ Rev.\ D {\bf 81}, 103532 (2010)
  [arXiv:1001.2301].
   
 \bibitem{Gumrukcuoglu:2010yc} 
  A.~E.~Gumrukcuoglu, B.~Himmetoglu, and M.~Peloso,
  Scalar-scalar, scalar-tensor, and tensor-tensor correlators from anisotropic inflation,
  Phys.\ Rev.\ D {\bf 81}, 063528 (2010)
  [arXiv:1001.4088].
  
\bibitem{Watanabe:2010bu} 
  M.~a.~Watanabe, S.~Kanno, and J.~Soda,
  Imprints of anisotropic inflation on the cosmic microwave background,
  Mon.\ Not.\ Roy.\ Astron.\ Soc.\  {\bf 412}, L83 (2011)
  [arXiv:1011.3604].
  
\bibitem{Bartolo:2012sd} 
  N.~Bartolo, S.~Matarrese, M.~Peloso, and A.~Ricciardone,
  Anisotropic power spectrum and bispectrum in the $f(\phi)F^2$ mechanism,
  Phys.\ Rev.\ D {\bf 87}, 023504 (2013)
  [arXiv:1210.3257].
  
		\bibitem{Ohashi:2013mka}
		J.~Ohashi, J.~Soda, and S.~Tsujikawa,
		Anisotropic non-Gaussianity from a two-form field,
		Phys. Rev. D \textbf{87}, 083520 (2013)
		[arXiv:1303.7340].
		
		\bibitem{Ohashi:2013qba}
		J.~Ohashi, J.~Soda, and S.~Tsujikawa,
		Observational signatures of anisotropic inflationary models,
		J. Cosmol. Astropart. Phys. \textbf{12}, 009 (2013)
		[arXiv:1308.4488].


\bibitem{Chen:2014eua} 
  X.~Chen, R.~Emami, H.~Firouzjahi, and Y.~Wang,
  The TT, TB, EB and BB correlations in anisotropic inflation,
  J. Cosmol. Astropart. Phys. {\bf 08} (2014) 027 
  [arXiv:1404.4083].


\bibitem{Soda:2012zm}
J.~Soda,
Statistical anisotropy from anisotropic inflation,
Class. Quant. Grav. \textbf{29}, 083001 (2012)
[arXiv:1201.6434].
	
		\bibitem{Maleknejad:2012fw}
A.~Maleknejad, M.~M.~Sheikh-Jabbari, and J.~Soda,
Gauge fields and inflation,
Phys. Rept. \textbf{528}, 161 (2013)
[arXiv:1212.2921].

	
		\bibitem{Do:2011zz}
		T.~Q.~Do and W.~F.~Kao,
		Anisotropic power-law inflation for the Dirac-Born-Infeld theory,
		Phys. Rev. D \textbf{84}, 123009 (2011).
		
		\bibitem{Ohashi:2013pca}
		J.~Ohashi, J.~Soda, and S.~Tsujikawa,
		Anisotropic power-law k-inflation,
		Phys. Rev. D \textbf{88}, 103517 (2013)
		[arXiv:1310.3053].
		
		\bibitem{Holland:2017cza}
		J.~Holland, S.~Kanno, and I.~Zavala,
		Anisotropic inflation with derivative couplings,
		Phys. Rev. D \textbf{97}, 103534 (2018)
		[arXiv:1711.07450].
		

		
		\bibitem{Nguyen:2021emx}
		D.~H.~Nguyen, T.~M.~Pham, and T.~Q.~Do,
		Anisotropic constant-roll inflation for the Dirac\textendash{}Born\textendash{}Infeld model,
		Eur. Phys. J. C \textbf{81}, 839 (2021)
		[arXiv:2107.14115].
		
		\bibitem{Do:2016ofi}
		T.~Q.~Do and W.~F.~Kao,
		Anisotropic power-law solutions for a supersymmetry Dirac\textendash{}Born\textendash{}Infeld theory,
		Class. Quant. Grav. \textbf{33}, 085009 (2016).
		
		\bibitem{Do:2020hjf}
		T.~Q.~Do,
		Stable small spatial hairs in a power-law k-inflation model,
		Eur. Phys. J. C \textbf{81}, 77 (2021)
		[arXiv:2007.04867].		
		
		
\bibitem{Pham:2023evo}
T.~M.~Pham, D.~H.~Nguyen, T.~Q.~Do, and W.~F.~Kao,
Anisotropic power-law inflation for models of non-canonical scalar fields non-minimally coupled to a two-form field,
Eur. Phys. J. C \textbf{84},  105 (2024)
[arXiv:2309.02690].
		
		
		\bibitem{Silverstein:2003hf}
		E.~Silverstein and D.~Tong,
		Scalar speed limits and cosmology: Acceleration from D-cceleration,
		Phys. Rev. D \textbf{70}, 103505 (2004)
		[hep-th/0310221].
		\bibitem{Alishahiha:2004eh}
		M.~Alishahiha, E.~Silverstein, and D.~Tong,
		DBI in the sky: Non-Gaussianity from inflation with a speed limit,
		Phys. Rev. D \textbf{70}, 123505 (2004)
		[hep-th/0404084].
		
		\bibitem{Armendariz-Picon:1999hyi}
		C.~Armendariz-Picon, T.~Damour, and V.~F.~Mukhanov,
		k-inflation,
		Phys. Lett. B \textbf{458}, 209 (1999)
		[hep-th/9904075].
		
		\bibitem{Garriga:1999vw}
		J.~Garriga and V.~F.~Mukhanov,
		Perturbations in k-inflation,
		Phys. Lett. B \textbf{458}, 219 (1999)
		[hep-th/9904176].

		\bibitem{Do:2020ler}
		T.~Q.~Do, W.~F.~Kao, and I.~C.~Lin,
		CMB imprints of non-canonical anisotropic inflation,
		Eur. Phys. J. C \textbf{81}, 390 (2021)
		[arXiv:2003.04266].
		
								
\bibitem{Saez:1986dil}
D.~S\'aez and V.~J.~Ballester,
A simple coupling with cosmological implications,
Phys. Lett. A \textbf{113}, 467 (1986).

\bibitem{Saez:1987st}
D.~S\'aez,
Simple coupling with cosmological implications. The initial singularity and the inflationary universe,
Phys. Rev. D \textbf{35}, 2027 (1987).

\bibitem{Singh:1991vou}
T.~Singh and A.~K.~Agrawal,
Some Bianchi-type cosmological models in a new scalar-tensor theory,
Astrophys. Space Sci. \textbf{182}, 289 (1991).

\bibitem{Singh}
 S. Ram and  J. K. Singh,  Cosmological models in certain scalar-tensor theories. Astrophys. Space Sci. {\bf 234}, 325 (1995).
 
\bibitem{Socorro:2009pt}
J.~Socorro, M.~Sabido, and L.~A.~Urena-Lopez,
Classical and quantum Cosmology of the Saez-Ballester theory,
Fizika B \textbf{19}, 177 (2010)
[arXiv:0904.0422].

\bibitem{Rasouli:2017glb}
S.~M.~M.~Rasouli and P.~Vargas Moniz,
Modified Saez\textendash{}Ballester scalar\textendash{}tensor theory from 5D space-time,
Class. Quant. Grav. \textbf{35}, 025004 (2018)
[arXiv:1712.01962].

\bibitem{Rasouli:2022hnp}
S.~M.~M.~Rasouli,
Noncommutativity, S\'aez\textendash{}Ballester theory and kinetic inflation,
Universe \textbf{8},  165 (2022)
[arXiv:2203.00765].

\bibitem{Quiros:2022vhm}
I.~Quiros and F.~A.~Horta-Rangel,
On the equivalence between S\'aez\textendash{}Ballester theory and Einstein-scalar field system,
Int. J. Mod. Phys. D \textbf{32},  2350033 (2023)
[arXiv:2209.00157].

\bibitem{Rasouli:2022tjn}
S.~M.~M.~Rasouli, M.~Sakellariadou, and P.~Vargas Moniz,
Geodesic deviation in S\'aez\textendash{}Ballester theory,
Phys. Dark Univ. \textbf{37}, 101112 (2022)
[arXiv:2203.00766].

\bibitem{Rasouli:2019axn}
S.~M.~M.~Rasouli, R.~Pacheco, M.~Sakellariadou, and P.~V.~Moniz,
Late time cosmic acceleration in modified S\'aez\textendash{}Ballester theory,
Phys. Dark Univ. \textbf{27}, 100446 (2020)
[arXiv:1911.03901].

\bibitem{Singh:2024zvm}
J.~K.~Singh, H.~Balhara, Shaily, T.~Q.~Do, and J.~Jena,
Observational constraints on Hubble parameter in S\'aez Ballester theory,
Astron. Comput. \textbf{47}, 100800 (2024).


\bibitem{Turner:1987bw}
M.~S.~Turner and L.~M.~Widrow,
Inflation produced, large scale magnetic fields,
Phys. Rev. D \textbf{37}, 2743 (1988).

\bibitem{Ratra:1991bn}
B.~Ratra,
Cosmological 'seed' magnetic field from inflation,
Astrophys. J. Lett. \textbf{391}, L1 (1992).

\bibitem{Bamba:2006ga}
K.~Bamba and M.~Sasaki,
Large-scale magnetic fields in the inflationary universe,
 J. Cosmol. Astropart. Phys. \textbf{02}, 030 (2007)
[astro-ph/0611701].

\bibitem{Martin:2007ue}
J.~Martin and J.~Yokoyama,
Generation of large-scale magnetic fields in single-field inflation,
J. Cosmol. Astropart. Phys. \textbf{01}, 025 (2008)
[arXiv:0711.4307].


\bibitem{Bamba:2008ja}
K.~Bamba and S.~D.~Odintsov,
Inflation and late-time cosmic acceleration in non-minimal Maxwell-$F(R)$ gravity and the generation of large-scale magnetic fields,
 J. Cosmol. Astropart. Phys. \textbf{04}, 024 (2008)
[arXiv:0801.0954].

\bibitem{Adak:2016led}
M.~Adak, \"O.~Akarsu, T.~Dereli, and \"O.~Sert,
Anisotropic inflation with a non-minimally coupled electromagnetic field to gravity,
 J. Cosmol. Astropart. Phys. \textbf{11}, 026 (2017)
[arXiv:1611.03393].

\bibitem{Do:2017onf}
T.~Q.~Do and W.~F.~Kao,
Anisotropic power-law inflation for a conformal-violating Maxwell model,
Eur. Phys. J. C \textbf{78},  360 (2018)
[arXiv:1712.03755].

\bibitem{Bamba:2012cp}
K.~Bamba, S.~Capozziello, S.~Nojiri, and S.~D.~Odintsov,
Dark energy cosmology: the equivalent description via different theoretical models and cosmography tests,
Astrophys. Space Sci. \textbf{342}, 155 (2012)
[arXiv:1205.3421].

\bibitem{Nojiri:2005pu}
S.~Nojiri and S.~D.~Odintsov,
Unifying phantom inflation with late-time acceleration: Scalar phantom-non-phantom transition model and generalized holographic dark energy,
Gen. Rel. Grav. \textbf{38}, 1285 (2006)
[arXiv:hep-th/0506212].

\bibitem{Capozziello:2005tf}
S.~Capozziello, S.~Nojiri, and S.~D.~Odintsov,
Unified phantom cosmology: Inflation, dark energy and dark matter under the same standard,
Phys. Lett. B \textbf{632}, 597 (2006)
[arXiv:hep-th/0507182].


\bibitem{Elizalde:2008yf}
E.~Elizalde, S.~Nojiri, S.~D.~Odintsov, D.~Saez-Gomez, and V.~Faraoni,
Reconstructing the universe history, from inflation to acceleration, with phantom and canonical scalar fields,
Phys. Rev. D \textbf{77}, 106005 (2008)
[arXiv:0803.1311].

		
\bibitem{Ackerman:2007nb} 
  L.~Ackerman, S.~M.~Carroll, and M.~B.~Wise,
  Imprints of a primordial preferred direction on the microwave background,
  Phys.\ Rev.\ D {\bf 75}, 083502 (2007)
 [astro-ph/0701357],
   [Erratum: Phys.\ Rev.\ D {\bf 80}, 069901(E)  (2009)].
  
  \bibitem{Groeneboom:2009cb} 
  N.~E.~Groeneboom, L.~Ackerman, I.~K.~Wehus, and H.~K.~Eriksen,
  Bayesian analysis of an anisotropic universe model: systematics and polarization,
  Astrophys.\ J.\  {\bf 722}, 452 (2010)
  [arXiv:0911.0150].
  
  \bibitem{Kim:2013gka} 
  J.~Kim and E.~Komatsu,
  Limits on anisotropic inflation from the Planck data,
  Phys.\ Rev.\ D {\bf 88}, 101301(R) (2013)
  [arXiv:1310.1605].


\bibitem{Ramazanov:2013wea} 
  S.~R.~Ramazanov and G.~Rubtsov,
  Constraining anisotropic models of the early Universe with WMAP9 data,
  Phys.\ Rev.\ D {\bf 89},  043517 (2014)
  [arXiv:1311.3272].

\bibitem{Ramazanov:2016gjl} 
  S.~Ramazanov, G.~Rubtsov, M.~Thorsrud, and F.~R.~Urban,
  General quadrupolar statistical anisotropy: Planck limits,
  J. Cosmol. Astropart. Phys. {\bf 03} (2017) 039 
  [arXiv:1612.02347].


\bibitem{Sugiyama:2017ggb} 
  N.~S.~Sugiyama, M.~Shiraishi, and T.~Okumura,
  Limits on statistical anisotropy from BOSS DR12 galaxies using bipolar spherical harmonics,
  Mon.\ Not.\ Roy.\ Astron.\ Soc.\  {\bf 473},  2737 (2018)
  [arXiv:1704.02868].
  
\bibitem{Do:2021pqk}
T.~Q.~Do and W.~F.~Kao,
Anisotropic hyperbolic inflation for a model of two scalar and two vector fields,
Eur. Phys. J. C \textbf{82}, 123 (2022)
[arXiv:2110.13516].

\bibitem{Bahamonde:2017ize}
S.~Bahamonde, C.~G.~B\"ohmer, S.~Carloni, E.~J.~Copeland, W.~Fang, and N.~Tamanini,
Dynamical systems applied to cosmology: dark energy and modified gravity,
Phys. Rept. \textbf{775-777}, 1 (2018)
[arXiv:1712.03107].

\end{thebibliography}
\end{document}